\documentclass[12pt]{article}
\usepackage{amsmath}
\usepackage{amssymb}
\usepackage{graphicx}
\usepackage{latexsym}
\usepackage{amssymb}

\newtheorem{definition}{Definition}
\newtheorem{theorem}{Theorem}
\newtheorem{proposition}{Proposition}

\newtheorem{corollary}{Corollary}
\newtheorem{example}{\it{Example}}
\newtheorem{remark}{{Remark}}
\newtheorem{Buchberger}{\bf Buchberger's algorithm}

\newtheorem{encoding}{\bf Systematic encoding algorithm}

\newtheorem{echelon}{\bf Echelon canonical form algorithm}

\newtheorem{Lally}{\bf Transpose algorithm}

\title{Computation of Gr\"obner basis for systematic encoding of generalized quasi-cyclic codes}
\author{
\large
Vo Tam Van\qquad Hajime Matsui\qquad Seiichi Mita\\
\normalsize
Dept.\ Electronics and Information Science\\
\normalsize
Toyota Technological Institute\\
\normalsize
2--12--1 Hisakata, Tenpaku, Nagoya, 468--8511, Japan}

\date{}

\begin{document}
\maketitle
\begin{abstract}
Generalized quasi-cyclic (GQC) codes form a wide and useful class of linear codes that includes thoroughly quasi-cyclic codes, finite geometry (FG) low density parity check (LDPC) codes, and Hermitian codes.
Although it is known that the systematic encoding of GQC codes is equivalent to the division algorithm in the theory of Gr\"obner basis of modules, there has been no algorithm that computes Gr\"obner basis for all types of GQC codes.
In this paper, we propose two algorithms to compute Gr\"obner basis for GQC codes from their parity check matrices: echelon canonical form algorithm and transpose algorithm.
Both algorithms require sufficiently small number of finite-field operations with the order of the third power of code-length.
Each algorithm has its own characteristic; the first algorithm is composed of elementary methods, and the second algorithm is based on a novel formula and is faster than the first one for high-rate codes.
Moreover, we show that a serial-in serial-out encoder architecture for FG LDPC codes is composed of linear feedback shift registers with the size of the linear order of code-length;
to encode a binary codeword of length $n$, it takes less than $2n$ adder and $2n$ memory elements.\\

\end{abstract}

\section{Introduction}\label{introduction}
Low density parity check (LDPC) codes were first discovered by Gallager \cite{Gallager} in 1962 and have recently been rediscovered and generalized by MacKay \cite{Mackay} in 1999.
The methods of constructing LDPC codes can be divided into two classes: random construction and algebraic one.

Random constructions of irregular LDPC codes \cite{Chung}\cite{Mackay}\cite{Shokrollahi} have shown the performance near to the Shannon limit for long code lengths of more than $10^7$ bits.
On the encoding of random LDPC codes, Richardson \emph{et al.} \cite{Richardson} proposed an efficient encoding method with the decomposition of the generator matrix into low triangular matrices, which was improved by Kaji \cite{Kaji} and Maehata \emph{et al.} \cite{Maehata} with another triangular (or LU-) factorization.
Both methods of encoding are based on the matrix multiplication.

\begin{figure}[t]
\centering
  \resizebox{10cm}{!}{\includegraphics{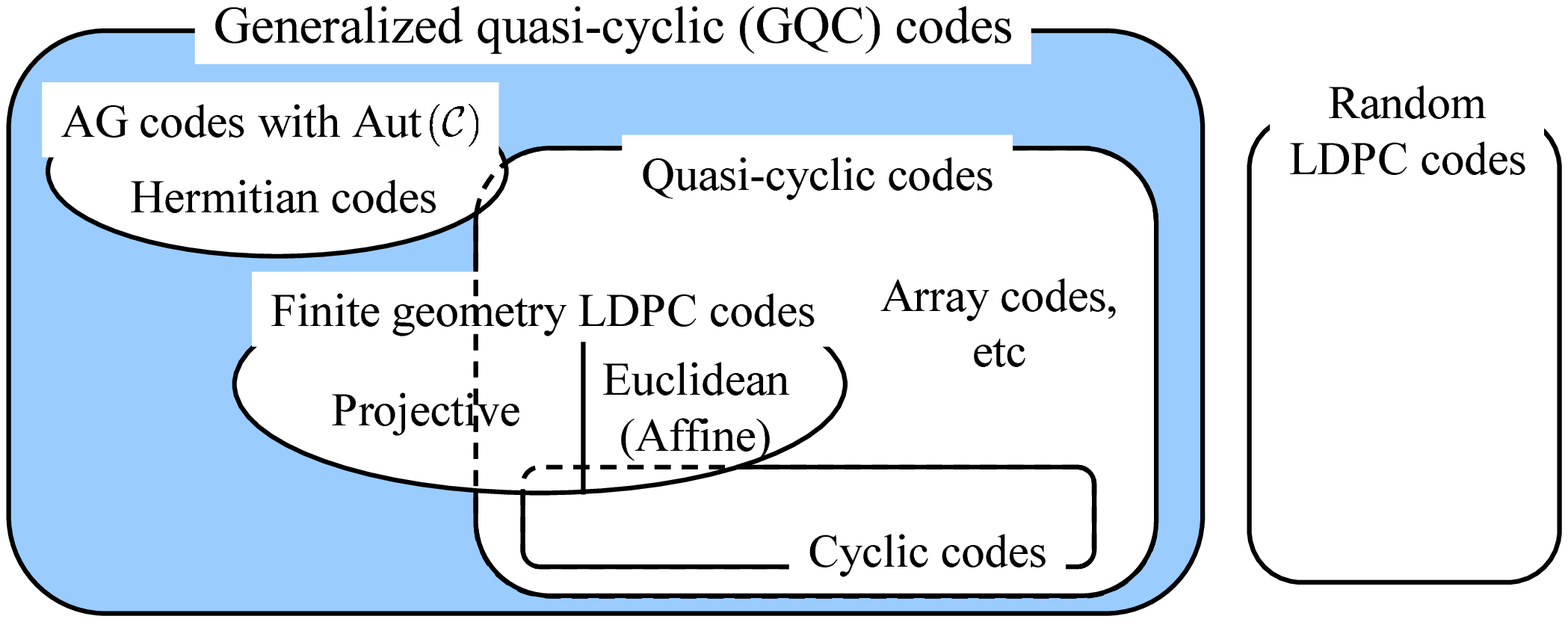}}
\caption{Inclusion-exclusion relation of various linear codes.\label{Venn}}
\end{figure}

There are many algebraic constructions of LDPC codes \cite{Andrews}\cite{Fan}\cite{Fossorier}\cite{Kamiya}\cite{Tanner}, which belong to a class of {\it quasi-cyclic (QC) codes} and provide efficient decoding performance.
Another remarkable algebraic construction of LDPC codes is finite geometry (FG) codes \cite{Kou}\cite{Lin}; These codes are divided into Euclidean (or affine) geometry (EG) codes, which are included in QC codes, and projective geometry (PG) codes, which are included not in QC codes but in broader {\it generalized quasi-cyclic (GQC) codes} (cf. Figure \ref{Venn}).
It can be stated briefly that GQC codes increase the randomness for QC codes and vary each length of cyclic parts in QC codes.

For several classes of QC LDPC codes, Fujita \emph{et al.} \cite{Fujita} proposed efficient encoding with circulant matrices and division technique.
With regard to GQC codes, which includes the algebraic LDPC codes, Heegard \emph{et al.} \cite{Heegard} showed that the systematic encoding was equivalent to the division algorithm of Gr\"obner bases, which generalize the generator polynomials in cyclic codes.
According to this work, Chen \emph{et al.} \cite{Chen} constructed an encoder architecture.
Thus, the encoding problem for GQC codes was changed into the computation of Gr\"obner basis.
For the computation of Gr\"obner basis for encoding GQC codes, Little \cite{Little} provided an algorithm for Hermitian codes, and Lally \emph{et al.} \cite{Lally} provided an algorithm for QC codes.
However, there has been no algorithm applicable to all GQC codes.

In this paper, we propose two algorithms for computing the Gr\"obner bases, which encode GQC codes, from their parity check matrices.
The first algorithm is based on Gaussian elimination, and the second algorithm is the generalization of Lally \emph{et al.}'s algorithm.
Both algorithms employ Buchberger's algorithm to create a Gr\"obner basis from codewords.
Moreover, in order to show its efficiency, we prove that the number of circuit elements in the encoder architecture is proportional to code-length for finite geometry codes.

A part of the first proposed algorithm to compute Gr\"obner basis was already known to some specialists in coding theory.
Kamiya \emph{et al.} \cite{Kamiya} announced that an encoder was obtained with fundamental row operation for a QC LDPC code from Euclidean geometry.
Recently, Little \cite{Little 2007} announced a similar result for a Hermitian code.
Our object is to provide algorithms computing the Gr\"obner bases for all GQC codes even in the case requiring column permutation.
On the other hand, the second proposed algorithm is based on a novel formula that produces Gr\"obner basis from that of the dual code.
The special case of our formula was found by Lally \emph{et al.} \cite{Lally} for QC codes.
In order to extend it to the case of GQC codes, we provide our formula with a completely different proof from that of Lally \emph{et al.}'s formula.
Both algorithms have $\mathcal{O}(n^3)$ order of the computational complexity, where $n$ is the code-length, and in fact for high-rate codes, we can show that the second has less complexity than the first.

Although the size of the encoder architecture for general GQC codes exceeds the linear order of code-length because of the number of orbits (cyclic parts), Chen \emph{et al.} \cite{Chen} proved that it had the linear order for Hermitian codes.
We newly prove that it also has the linear order for FG codes.
While Richardson \emph{et al.}'s and Kaji's methods for general LDPC codes run by the linear order of finite-field {\it operations}, our encoder architecture for FG codes can achieve not only the linear order of operations but also the linear order of {\it circuit elements} and no latency.
In addition, our encoder architecture for the binary FG LDPC codes requires only adder elements without multiplication (i.e., no AND element).

This paper deals with all GQC codes; Siap \emph{et al.} \cite{Siap} mainly focused on one-generator GQC codes.
Another example of GQC codes is the class of algebraic geometry (AG) codes with automorphism groups \cite{Heegard}, including Hermitian codes \cite{Little}.
It is worthy to notice that GQC codes include two remarkable classes of Hermitian codes and some PG codes outside QC codes.
Thus, GQC codes form the vastest algebraic class in linear codes that holds compact encoder architecture.
Therefore, we can choose more appropriate and high-performance codes from GQC codes than those from QC codes.

This paper is organized as follows.
Section 2 provides the definition of GQC codes and the techniques of Gr\"obner basis.
Section 3 provides the details of the first: echelon canonical form algorithm.
Section 4 provides the details of the second: transpose algorithm.
In section 5, we estimate the computational complexity of proposed two algorithms.
In section 6, we prove the linearity of the circuit-scale of the encoder architecture for FG LDPC codes.
Finally, we conclude this paper in section 7.

\section{Preliminaries}\label{preliminaries}
Throughout the paper, we denote $A:=B$ if $A$ is defined as $B$.
First, we describe the definition and module structure of generalized quasi-cyclic codes.
Then, we review Gr\"obner basis of modules over polynomial rings; the complete theory of Gr\"obner basis is referred to \cite{Becker}\cite{Cox}, and that of automorphism group and orbit is referred to \cite{MacWilliams}.

\subsection{Definitions}
Consider a linear code $\mathcal{C}\subset\mathbb{F}_q^{\,n}$ of length $n$, where $q$ is a prime power and $\mathbb{F}_q$ is $q$-element finite field.
Let $S$ be the set of locations (that is, coordinate positions) of codewords in $\mathcal{C}$: $\mathcal{C}\ni c=(c_s)_{s\in S}$.
Without loss of generality, we set $S=\{1,2,\cdots,n\}$.
Now suppose that there is a decomposition of $S$,
\begin{equation}\label{orbit}
S=\bigcup_{i=1}^mO_i\,,\quad
|S|=n=\sum_{i=1}^{m}l_i,\quad l_i:=|O_i|,
\end{equation}
and accordingly, decompose any codeword $c\in \mathcal{C}$ into $m$ shortened codes:
\begin{equation}\label{decomposition}
c=(c_1,c_2,\cdots,c_m),
\end{equation}
where $c_i$ is a shortened codeword dropping components outside $O_i\,$.
Consider simultaneous local cyclic shift $\sigma$ of each $c_i$ satisfying
\begin{align}
\sigma(c)&:=(\sigma(c_1),\cdots,\sigma(c_m)),\nonumber\\
\sigma(c_i)&:=(c_{i,l_i-1},c_{i,0},\cdots,c_{i,l_i-2})\label{sigma}\\
\mathrm{for}\;
c_i&=(c_{i,0},c_{i,1},\cdots,c_{i,l_i-1}).\nonumber
\end{align}

\begin{definition}\label{cyclical}
If we have $m<n$ and $\sigma(c)\in \mathcal{C}$ for all $c\in \mathcal{C}$,
then we call a pair of $\mathcal{C}$ and $\sigma$ {\it a generalized quasi-cyclic code (GQC code)}.
\hfill$\Box$
\end{definition}

If $\mathcal{C}$ is GQC, thus we obtain a nontrivial $\sigma$ of the automorphism group Aut$(\mathcal{C})$ of $\mathcal{C}$.
Conversely, if Aut$(\mathcal{C})$ includes $\sigma\not=1$,
then the cyclic group $\langle\sigma\rangle$ generated by $\sigma$ defines {\it orbit} $O(s):=\{\sigma^l(s)\,|\,\sigma^l\in\langle\sigma\rangle\}$ of $s\in S$.
Note that we have $O(s)=O(s')$ for $s'\in O(s)$, and that we have $O(s)=\{s\}$ if $\sigma(s)=s$.
Then, $S$ is equal to the disjoint union of distinct orbits as described in \eqref{orbit}, where $O_i:=O(s_i)$ for some $s_i\in S$.
We can regard $\sigma$ as the simultaneous shift \eqref{sigma} of $\{c_i\}$.
Thus, we have shown that the class of GQC codes agrees with the class of linear codes with nontrivial Aut$(\mathcal{C})\supset\langle\sigma\rangle$.

\begin{remark}\label{product}
{\rm
Thus, we see that each $\mathcal{C}_i:=\{c_i\}$ decides a cyclic code. However, in general the whole $\mathcal{C}$ does not agree with the combined code $\prod \mathcal{C}_i\ni[c_1\cdots c_m]$, since the individual shift, for example $(\sigma(c_1),c_2,\cdots,c_m)$, does not generally belong to $\mathcal{C}$. We will see the difference between $\mathcal{C}$ and $\prod \mathcal{C}_i$ at \eqref{polynomial matrix} in subsection \ref{Grobner basis of module}.
}
\end{remark}
\begin{remark}
{\rm
Siap \emph{et al.} \cite{Siap} define GQC codes as $\mathbb{F}_q[t]$-submodules of a certain module $M$ at \eqref{module} in the next subsection, where we will see that their definition of GQC codes is equivalent to our definition.}
\end{remark}

Usually, the generator matrix of a linear code indicates the matrix whose rows are linearly independent and compose a basis of the linear space. We often relax this definition for convenience; we call a generator matrix of a linear code the matrix whose rows are not too many and contain the basis.

\begin{example}\label{ex:generator}
{\rm
Consider the linear code $\mathcal{C}_{1}\subset\mathbb{F}_2^{\,7}$ defined by a generator matrix as below.
\begin{center}
$G_1 = $
$\left(
\begin{tabular}{ccc|ccc|c}
  1& 1& 1& 0& 0& 0 & 1 \\
\hline
  1& 1& 0& 1& 0& 1 & 0\\
  0& 1& 1& 1& 1& 0 & 0\\
  1& 0& 1& 0& 1& 1 & 0
\end{tabular}
\right)$
\end{center}
Since the second row plus the third row equals the fourth row, we see that the dimension of $\mathcal{C}_1$ is three.
If we apply the permutation $\sigma$ given by \eqref{sigma}, then a codeword in $\mathcal{C}_1$
is transferred into another codeword in $\mathcal{C}_1$.
Thus, $\mathcal{C}_1$ is made from 4 cyclic codes defined by
$(1)$, $(1\,1\,1)$,
$\left(\!
\begin{tabular}{c}
 1 1 0 \\
 0 1 1 \\
 1 0 1
\end{tabular}
\!\right)$, and $\left(\!
\begin{tabular}{c}
 1 0 1 \\
 1 1 0 \\
 0 1 1
\end{tabular}
\!\right)$ (and two all-zero codes),
and $\mathcal{C}_1$ is a GQC code with 3 orbits.
}
\hfill$\Box$
\end{example}

\begin{figure}[h!]
\begin{center}
  \resizebox{5.5cm}{!}{\includegraphics{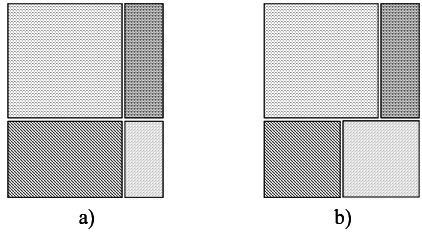}}
  \caption{Intuitive models of generator matrices made from four matrices. Model a) defines a 2-orbit GQC code, but Model b) does not define a GQC code.\label{subcodes}}
\end{center}
\end{figure}

Note that, if $l_1  = l_2 = \cdots =l_m$, then $\mathcal{C}$ is a quasi-cyclic code \cite{Lally}\cite{Lin}\cite{Peterson}.
Moreover, if $m=1$, then we come back to cyclic code.
In order to increase the randomness of the codes, it is desirable that we can combine various circulant matrices (cf. Section \ref{generalized}) to generate new GQC codes.
The code in Figure \ref{subcodes}, a) shows us a 2-orbit GQC code constructed from four matrices.
On the other hand, Figure \ref{subcodes}, b) shows that the code is also obtained from four matrices but is not a GQC code.

\subsection{Module structure of generalized quasi-cyclic codes}
Let $\mathcal{C}$ be a GQC code with a permutation $\sigma$.
Under the action of $\left \langle \sigma \right \rangle$,
we can decompose $c \in \mathcal{C}$ into $m$ shortened codes as described in \eqref{decomposition}.
Pick $c_i$ and label it as $c_i = (c_{i,j})$
where $j = 0,\cdots,l_i  - 1$ with $l_i:=\left|{O_{i} }\right|$.
For convenience, we decide that the second index is an integer modulo $l_i$,
and the permutation $\sigma$ of \eqref{sigma} means
$\sigma(c_{i,j})=c_{i,(j-1 \,\mathrm{mod}\, l_i)}$  for all $i = 1,\cdots,m$ and $j = 0,\cdots,l_i-1$.
Then, a codeword in $\mathcal{C}$ can be represented as an $m$-tuple of polynomials in $\mathbb{F}_q[t]$:
$$c = (c_1 (t),c_2 (t),\cdots,c_m (t)),$$
where $c_i (t) = \sum\limits_{j = 0}^{l_i  - 1} {c_{i,j} t^j } $.
Thus, $\mathcal{C}$ is regarded as
a linear subspace of $M$, where
\begin{equation}\label{module}
    M:=\bigoplus _{i = 1}^m \Big(\mathbb{F}_q[t]/\left(t^{l_i }-1\right)\Big)
\end{equation}
and $\mathbb{F}_q[t]/\left(t^{l_i }-1\right)$ is the quotient ring by an ideal $\left(t^{l_i }-1\right):=\left(t^{l_i }-1\right)\mathbb{F}_q[t]$.
Moreover, we can regard the action of $\sigma$ as the multiplication of $t$ as follows:
\begin{equation}
tc_i(t)=\!\sum\limits_{j=0}^{l_i-1}c_{i,j}t^{j + 1}\equiv\!\sum\limits_{j=0}^{l_i-1}c_{i,j-1}t^j=\!\sum\limits_{j=0}^{l_i-1}{\sigma(c_{i,j})t^j},
\end{equation}
where ``$\equiv$'' means the equality modulo $\left(t^{l_i }-1\right)$.
We can see that multiplying $c$ by $t$ is equivalent to permuting the
codeword locally cyclically by $\sigma$.
Thus, $\mathcal{C}$ is closed under the multiplication by $t$
and $\mathcal{C}$ is considered as an $\mathbb{F}_q[t]$-submodule of $M$.
For convenience to compute Gr\"obner basis, we consider the following natural map: $\pi :\mathbb{F}_q[t]^m  \to  M$.
Let $e_i$ be the $i$-th standard basis vector in $\mathbb{F}_q[t]$-module $\mathbb{F}_q[t]^m$, that is,
\begin{align*}
e_1:=(1,0,0,\cdots,0),\;
e_2&:=(0,1,0,\cdots,0),\\
&\cdots,\;e_m:=(0,0,0,\cdots,1),
\end{align*}
and $X_i:=\left(t^{l_i }  - 1\right)e_i$ for $i = 1,\cdots,m$.
Define $\overline{\mathcal{C}}:= \pi ^{ - 1} (\mathcal{C})$, which is a submodule of $\mathbb{F}_q[t]^m$ and is generated by all codewords
in $\mathcal{C}$ (regarded as vectors in $\mathbb{F}_q[t]^m$) and all  $X_i$'s, that is,
\begin{equation}\label{submodule}
\overline{\mathcal{C}}= \mathcal{C} + \left\langle{X_i \;|\; i=1,\cdots,m}\right\rangle,
\end{equation}
where $\left\langle{X_i \;|\; i=1,\cdots,m}\right\rangle$ indicates the submodule generated by all $X_i$.

\subsection{Gr\"obner basis of $\mathbb{F}_q[t]$-module
\label{Grobner basis of module}}
We call an element of the form $t^j e_i$ a {\it monomial} in $\mathbb{F}_q[t]^m$.
Then, any polynomial vector in $\mathbb{F}_q[t]^m$ can be represented as a linear combination of monomials.

Although {\it Gr\"obner basis} of a submodule in $\mathbb{F}_q[t]^m$ is determined for each {\it monomial ordering}, only the following two orderings are required in this paper.
The {\it position over term (POT)} ordering \cite{Heegard} on $\mathbb{F}_q [t]^m $ is defined by $t^l e_i  > _\mathrm{POT} t^k e_j $ if $i < j$, or $i = j$ and $l > k$.
Then, we have
$
  e_1  >_\mathrm{POT} e_2  >_\mathrm{POT} \cdots >_\mathrm{POT} e_m
$.

Similarly, the {\it reverse POT (rPOT)} ordering is defined by $t^l e_i  > _\mathrm{rPOT} t^k e_j $ if $i > j$, or $i = j$ and $l > k$.
Then, we have
$
   e_1  <_\mathrm{rPOT} e_2  <_\mathrm{rPOT} \cdots <_\mathrm{rPOT} e_m.
$

For a polynomial $f(t):=\sum_{i=0}^{d}f_it^i\in\mathbb{F}_q[t]$ with $f_d\not=0$, we define {\it degree} of $f(t)$ by $\deg(f):=d$, and we say that $f$ is monic if $f_d=1$.
Thus, we can define two types of Gr\"obner bases for the submodule $\overline{\mathcal{C}}$ of $\mathbb{F}_q[t]^m$ associated with an $m$-orbit GQC code $\mathcal{C}$.

\begin{definition}\label{Grobner}
We define {\it POT Gr\"obner basis} of $\overline{\mathcal{C}}$ as the following set $\mathcal{G}=\{g_1,g_2,\cdots,g_m\}$ of polynomial vectors
\begin{equation}\label{polynomial matrix}
\begin{array}{cclcccc}
 g_1&=&(g_{11} (t),&g_{12} (t),&\cdots ,&g_{1m} (t) ), \\
 g_2&=&(0,&g_{22} (t),&\cdots ,&g_{2m} (t) ), \\
\vdots& \empty &\;\;\vdots&\ddots& \ddots & \vdots \\
 g_m& =&(0,&\cdots,&0,&g_{mm} (t) )
\end{array}
\end{equation}
such that $g_1,\cdots,g_m\in\overline{\mathcal{C}}$ and $g_{ii}(t)$ has the minimum degree among the vectors of the form $(0,\cdots,0,c_i(t),\cdots,c_m(t))\in\overline{\mathcal{C}}$ with $c_i(t)\not=0$.
If $g_{ii}$'s are monic and $\mathcal{G}$ satisfies $\deg g_{ij}  < \deg g_{jj} $ for all $1 \le i < j\le m$, then we call it {\it reduced} POT Gr\"obner basis.
Moreover, we define {\it rPOT Gr\"obner basis} of $\overline{\mathcal{C}}$ as the following set $\mathcal{H}=\{h_1,h_2,\cdots,h_m\}$ of polynomial vectors
\begin{equation}\label{polynomial dual}
\hspace{-2mm}
\begin{array}{cclcccc}
h_1&\!\!\!\!=&\!\!\!(h_{11}(t),&\!\!\!0,&\!\!\!\cdots,&\hspace{-5mm}0),\\
\vdots&\!\!\!\! \empty &\!\!\!\;\;\vdots&\!\!\!\ddots&\!\!\! \ddots &\hspace{-5mm} \vdots \\
h_{m-1}&\!\!\!\!=&\!\!\!(h_{m-1,1}(t),&\!\!\!\cdots,&\!\!\!h_{m-1,m-1}(t),&\hspace{-5mm}0), \\
h_m&\!\!\!\!=&\!\!\!(h_{m1}(t),&\!\!\!\cdots,&\!\!\!h_{m,m-1}(t),&\hspace{-5mm}h_{mm}(t))
\end{array}
\end{equation}
such that $h_1,\cdots,h_m\in\overline{\mathcal{C}}$ and $h_{ii}(t)$ has the minimum degree among the vectors of the form $(c_1(t),\cdots,c_i(t),0,\cdots,0)\in\overline{\mathcal{C}}$ with $c_i(t)\not=0$.
If $h_{ii}$'s are monic and $\mathcal{H}$ satisfies $\deg h_{ij}  < \deg h_{jj} $ for all $1\le j < i \le m$, then we call it {\it reduced} rPOT Gr\"obner basis.
\hfill$\Box$
\end{definition}

Since $\left(t^{l_i}-1\right)e_i$'s are included in $\overline{\mathcal{C}}$, the diagonal polynomials $g_{ii}$ and $h_{ii}$ divide $t^{l_i}-1$, and both Gr\"obner bases of $\overline{\mathcal{C}}$ exist.
If $g_{ij}=0$ for all $i\not=j$, then $\mathcal{C}$ agrees with the combined code $\prod \mathcal{C}_i$ as noticed at Remark \ref{product}.

From any Gr\"obner basis, we can easily obtain the reduced Gr\"obner basis by fundamental row operations of polynomial matrix.
Each GQC code has its unique reduced Gr\"obner basis.

The Gr\"obner basis of $\overline{\mathcal{C}}$ has two important roles; one is that it generates $\mathcal{C}$, and the other is division algorithm (which is stated in the next subsection). Any element $c\in\overline{\mathcal{C}}$ has the following expression
\begin{equation}\label{generation}
   c = P_1 (t)g_1  + \cdots + P_m (t)g_m,
\end{equation}
where $P_i (t) \in \mathbb{F}_q[t]$.
If $\deg(P_ig_{ii})<l_i$ for all $i$, then we have $c\in \mathcal{C}$ strictly.
For the use of encoding, we define {\it redundant monomial} as $t^je_i$ with $0\le j<\deg g_{ii}(t)$ (standard monomial in \cite{Heegard}). The other types of monomial $t^je_i$ with $\deg g_{ii}(t)\le j<l_i$ are called {\it non-redundant} (or {\it information}) monomial.
It follows from \eqref{generation} that the number of information monomials equals the dimension of $\mathcal{C}$.

By minimizing $\deg g_{ii}$, we can obtain $\mathcal{G}$ from the generator matrix. This procedure is called {\it Buchberger's algorithm}, which is now described for our situation.

\begin{Buchberger}\hfill
{\rm
\\
\textbf{Input:} A $k\times n$ generator matrix $G$ of a GQC code $\mathcal{C}$\\
\textbf{Output:} A POT Gr\"obner basis
$\mathcal{G} = \left\{g_1,\cdots,g_m\right\}$.
\\
\textbf{Step 1.} Regard the row $G_i$ of $G$ as polynomial vector $G_i=(G_{i1},\cdots,G_{im})\in\mathbb{F}_q[t]^m$.\\
\textbf{Step 2.} For $j=1$ to $m$;
\begin{align*}
&\left[\!
\begin{array}{l}
\vspace{1mm}
g_{j}:=\sum_{i=j}^{k}Q_{ij}G_{i}\;\;\mbox{with}\;\{Q_{jj},\cdots,Q_{kj}\}\;\;\mbox{such}\;\mbox{that}\\
\vspace{1mm}
\quad g_{jj}:=\mathrm{gcd}\{G_{jj},\cdots,G_{kj}\}=\sum_{i=j}^{k}Q_{ij}G_{ij}\,.\\
\vspace{1mm}\mbox{For}\;\;l=j+1\;\;\mbox{to}\;\;k;\\
\quad\Big[\:G_l:=G_l-R_l(t)g_{j}\;\;\mbox{with}\;\;R_{l}:=G_{lj}/g_{jj}.
\end{array}\right.
\end{align*}
\textbf{Step 3.} If $g_j$ is zero vector, then put $g_j:=X_j$. Reduce $\mathcal{G}$ by fundamental row operations.
}
\hfill $\Box$
\end{Buchberger}

\begin{example}\label{monomials}
{\rm
Consider again \emph{Example} \ref{ex:generator}.
Buchberger's algorithm is applied to $G_1$.
We obtain $g_{1}=(1,1+t,1)$ by adding the third row to the first row.
Since the other polynomial vectors are the multiple of $g_1$, we obtain $g_2=X_2$ and $g_3=X_3$ as shown in Figure \ref{example}.
Then, the information monomials are $t^2e_1, te_1, e_1$
and the redundant monomials are $t^2e_2, te_2, e_2, e_3$.
\hfill$\Box$
\begin{figure}[h!]
\begin{center}
  \resizebox{8.3cm}{!}{\includegraphics{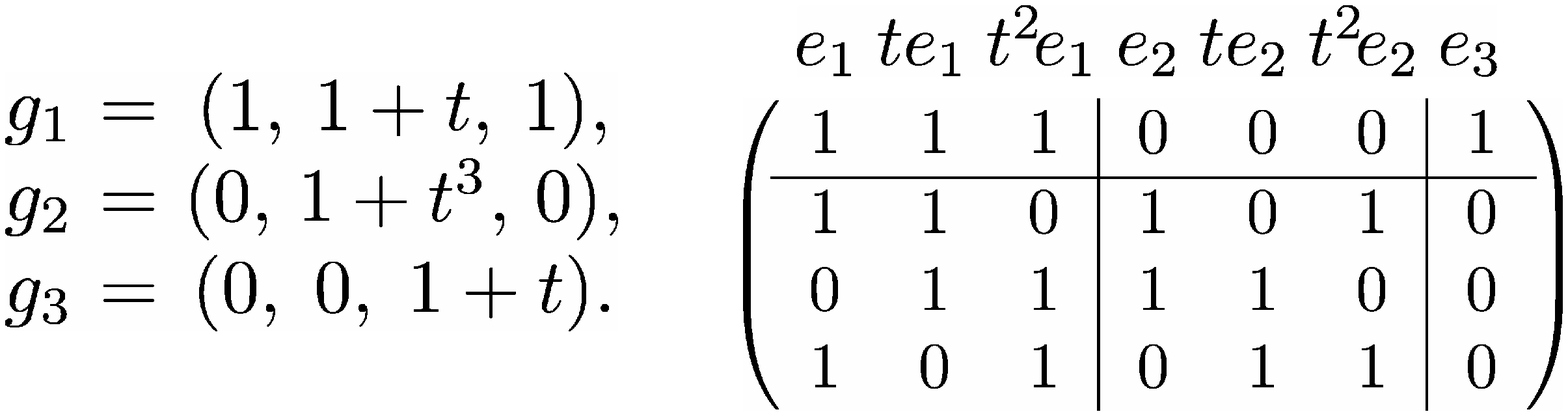}}
\caption{The reduced POT Gr\"obner basis of $\overline{\mathcal{C}}_1$ and monomials.\label{example}}
\end{center}
\end{figure}
}
\end{example}

\subsection{Systematic encoding algorithm}
Once a Gr\"obner basis $ \mathcal{G} = \{ g_1 ,\cdots,g_m \} $ of $\overline{\mathcal{C}}$ is obtained, then division algorithm with respect to $ \mathcal{G}$ can be applied to $u\in\mathbb{F}_q[t]^m $ to obtain the following representation
\begin{equation}\label{remainder}
   u = Q_1 (t)g_1  + \cdots + Q_m (t)g_m  + \overline{u},
\end{equation}
where $Q_i(t)\in\mathbb{F}_q[t]$, and $\overline{u}= (\overline{u}_1(t),\cdots,\overline{u}_m(t))$ with $\deg\overline{u}_i<\deg g_{ii}$.
In other words, $\overline{u}$ is a unique linear combination of redundant monomials.
It follows from \eqref{generation} and \eqref{remainder} that $u \in\overline{\mathcal{C}}\Leftrightarrow \overline{u}=\left(0,\cdots, 0\right)$, which generalizes the condition of codewords in cyclic codes.
Then, the encoding of $\mathcal{C}$ is described as follows.

\begin{encoding}\hfill
{\rm
\\
\textbf{Input:} Information symbols $u\in \mathbb{F}_q^k $ and Gr\"obner basis $\mathcal{G} = \left\{g_1,\cdots,g_m\right\}$.\\
\textbf{Output:} Encoded codeword $c \in \mathcal{C}$.
\\
\textbf{Step 1.} Calculate $u\in\mathbb{F}_q[t]^m$ as a linear combination of information symbols and information monomials.
\\
\textbf{Step 2.} Put $u_1=(u_{11}(t),\cdots,u_{1m}(t)):=u$;
\begin{align*}
&\mbox{For}\;i=1\;\mbox{to}\;m;\\
&\;\left[\!
\begin{array}{l}
\vspace{1mm}
\mbox{Find}\;\;Q_i(t)\;\;\mbox{and}\;\;\overline{u}_i(t)\;\;\mbox{such}\;\mbox{that}\\
\vspace{1mm}
\quad u_{ii}(t)=Q_i(t)g_{ii}(t)+\overline{u}_i(t),\;\;
\deg\overline{u}_i<\deg g_{ii}\,.\\
\vspace{1mm}
\mbox{Calculate}\;\;u_{i+1}:=u_i-Q_i(t)g_i\quad(\in M)\\
\quad=(\overline{u}_1(t),\cdots,\overline{u}_i(t),u_{i+1,i+1}(t),\cdots,u_{i+1,m}(t)).
\end{array}\right.
\end{align*}
Put $\overline{u}:=(\overline{u}_1(t),\cdots,\overline{u}_m(t))
=u-\sum_{i=1}^m Q_i(t)g_i$ in $M$.\\
\textbf{Step 3.} By subtraction $c:=u-\overline{u}$, we obtain the encoded codeword $c \in \mathcal{C}$.\\
}
\hfill $\Box$
\end{encoding}

Step 2 itself is called {\it division algorithm}, which generalizes the classical polynomial division in the encoding of cyclic codes.
Thus, another merit of considering the reduced Gr\"obner basis is that it reduces the computational complexity of the division algorithm.

\begin{example}
{\rm
We reuse the GQC code $\mathcal{C}_1$ with the reduced Gr\"obner basis $\mathcal{G}_{1}$.
The information symbols for $\mathcal{C}_1$ can be taken as the coefficients of information monomials $\{e_1,te_1,t^2e_1\}$.
We apply the systematic encoding algorithm
to encode message $(1,0,1)$.
First, we put $u:=e_1+0\cdot te_1+t^2e_1=(1+t^2,0,0)$.
Then, we divide $u$ by $\mathcal{G}_1$ to obtain the remainder (or parity symbols) $\overline{u}$:
$$
\begin{array}{lll}
\overline{u}=u - (1+t^2)g_1=(0,t+t^2,0),
\end{array}
$$
where the last equality follows from $(0,1+t^3,0)=(0,0,0)$ in $M$.
Since $\overline{u}$ contains only redundant monomials, we finish the division algorithm.
Thus, the encoded polynomial vector is
$ u- \overline{u} = (1+t^2,t+t^2,0)$ and the corresponding encoded codeword $c$ is
equal to $(1010110)$. Since we have
$$
\big(\!\!\begin{tabular}{c}
1 0 1 0 1 1 0
\end{tabular}\!\!\big)
=
\big(\!\!\begin{tabular}{c}
0 1 1 0
\end{tabular}\!\!\big)\!
\left(\!
\begin{tabular}{c|c|c}
  1 1 1& 0 0 0 & 1 \\
\hline
  1 1 0& 1 0 1 & 0\\
  0 1 1& 1 1 0 & 0\\
  1 0 1& 0 1 1 & 0
\end{tabular}
\!\right),
$$
we can check that $c \in\mathcal{C}_1$, which is required.
}
\hfill $\Box$
\end{example}

\section{Computing Gr\"obner basis from parity\\
check matrix with echelon canonical form}\label{echelon algorithm}
In this section, we consider the problem about computing Gr\"obner basis, which generates a GQC code, from a given parity check matrix.
In many situations, each GQC code $\mathcal{C}$ is specified by a parity check matrix, that is, the generator matrix of its dual code $\mathcal{C}^\bot$.
Since we have $\mathrm{Aut}(\mathcal{C})=\mathrm{Aut}(\mathcal{C}^\bot)$, both codes are viewed as the submodules of the same $M$ at \eqref{module}.

\begin{figure*}[t!]
\begin{center}
  \includegraphics[scale=0.7]{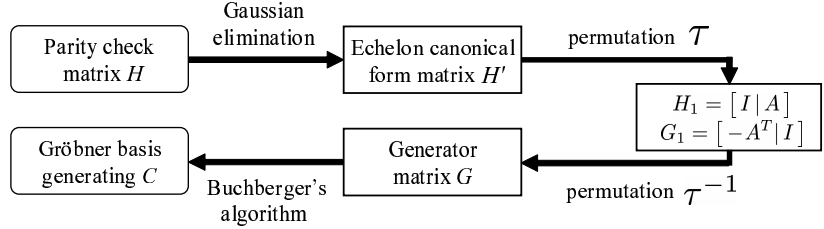}
\vspace{-0.5em}
  \caption{Outline of computing Gr\"obner basis for encoding by echelon canonical form algorithm from parity check matrix.
\label{echelon canonical}}
\end{center}
\end{figure*}

Before describing the proposed algorithm, we remind that elementary row operations can be used to simplify
a matrix and we obtain \emph{echelon\ canonical form} \cite{Peterson}, which is defined as follows:
\begin{itemize}
 \item   Every leftmost non-zero value is 1.
 \item   Every column containing the leftmost non-zero value has all zero other entries.
 \item   The leftmost non-zero value in any row is on the right of that in every preceding row.
\end{itemize}
For example, we consider the following two matrices:
$$
\left (
 \begin{tabular}{l}
 \underline{1} 0 1 0 0 0 0 1 \\
 0 \underline{1 1} 0 0 1 0 0 \\
 0 0 0 \underline{1} 0 0 0 1 \\
 0 0 0 0 \underline{1 1} 0 0 \\
 0 0 0 0 0 0 \underline{1 1}
 \end{tabular}
\right ),
\quad
\left (
\begin{tabular}{l}
 1 0 1 1 1 1 0 0 \\
 0 1 1 0 0 1 0 0 \\
 1 0 1 0 1 1 1 0 \\
 1 1 0 1 0 1 0 0 \\
 0 1 1 0 0 1 1 1
\end{tabular}
\right )
.
$$
The left matrix is the echelon canonical form of the right matrix.

By using the echelon canonical form, we can compute the Gr\"obner basis of $\overline{\mathcal{C}}$ from parity check matrix.
The flow of our first algorithm is presented in Figure \ref{echelon canonical} and described as follows:

\begin{echelon}\hfill
\\
{\rm
\textbf{Input:} Parity check matrix $H$ of a GQC code $\mathcal{C}$.
\\
\textbf{Output:} POT Gr\"obner basis $\mathcal{G}$ of $\overline{\mathcal{C}}$.
\\
\textbf{Step 1.} Transform $H$ to echelon canonical form $H'$ by Gaussian elimination.
\\
\textbf{Step 2.} Select permutation $\tau$ satisfying
$H_1:=\tau(H')=[I|A]$, and then $G_1 := [{-A}^T |I]$.
\\
\textbf{Step 3.} Compute generator matrix $G=\tau^{-1}(G_1)$.
\\
\textbf{Step 4.} Obtain $\mathcal{G}$ by Buchberger's algorithm from $G$.
}
\hfill $\Box$
\end{echelon}
In the step $2$ of the above algorithm, $G_1$ satisfies the equation $G_1 \times H_1^T = 0$ and  $\tau^{-1}(G_1) \times \{\tau^{-1}(H_1)\}^T=\tau^{-1}(G_1) \times {H'}^T = 0$.
Therefore, we permute the column vectors of $G_1$ by $\tau^{-1}$ to obtain a generator matrix $G$ of the GQC code $\mathcal{C}$.

\begin{example}
{\rm
Let $\mathcal{C}_2$ be a GQC code defined by the following parity check matrix $H_2$:
}
\end{example}
$$ H_2 := \left (
\begin{tabular}{c|c|c}
 1 1 0 1 1 0 & 1 0 1 1 0 0 & 1 1 0 \\
 0 1 1 0 1 1 & 0 1 0 1 1 0 & 0 1 1 \\
 1 0 1 1 0 1 & 0 0 1 0 1 1 & 1 0 1 \\
 1 1 0 1 1 0 & 1 0 0 1 0 1 & 1 1 0 \\
 0 1 1 0 1 1 & 1 1 0 0 1 0 & 0 1 1 \\
 1 0 1 1 0 1 & 0 1 1 0 0 1 & 1 0 1
\end{tabular}
\right ).$$
We can see that $\mathcal{C}_2$ has locally cyclic property with the column permutation
$\sigma=(1 \cdots 6)(7 \cdots 12)(13 \cdots 15)$,
where, e.g., $(1\cdots6)$ indicates permutation $1\rightarrow2\rightarrow\cdots\rightarrow6\rightarrow1$,
and $\mathcal{C}_2$ has 3 orbits: $l_1=l_2=6$ and $l_3=3$.
Firstly, we use Gaussian elimination to transform $H_2$ to the equivalent echelon canonical form
\begin{equation}\label{stair form}
 \left (
 \begin{tabular}{c|c|c|c}
 \underline{1} 0 & 1 1 0 1 & 0 0 0 0 & 1 0 1 0 1\\
 0 \underline{1} & \underline{1 0 1 1} & 0 0 0 0 & 1 1 0 1 1\\
 0 0 & 0 0 0 0 &\underline{1} 0 0 0 & 1 1 0 0 0\\
 0 0 & 0 0 0 0 & 0 \underline{1} 0 0 & 1 0 0 0 0\\
 0 0 & 0 0 0 0 & 0 0 \underline{1} 0 & 0 1 0 0 0\\
 0 0 & 0 0 0 0 & 0 0 0 \underline{1} & 1 1 0 0 0
\end{tabular}
\right ).
\end{equation}
If we choose the column permutation $\tau$ such that
the set of column location $\left( 1,2,\cdots,14,15  \right ) $ is mapped by $\tau$ to
$$\left( 1,2,\;\: 7,8,9,10, \;\: 3,4,5,6, \;\: 11,12,13,14,15 \right),$$
then matrix \eqref{stair form} is transformed to the standard form matrix $[I|A]$:
$$
 \left (
 \begin{tabular}{c|c|c|c}
 \underline{1} 0 &0 0 0 0& 1 1 0 1 & 1 0 1 0 1 \\
 0 \underline{1} &0 0 0 0& 1 0 1 1 & 1 1 0 1 1 \\
 0 0 &\underline{1} 0 0 0& 0 0 0 0 & 1 1 0 0 0 \\
 0 0 &0 \underline{1} 0 0& 0 0 0 0 & 1 0 0 0 0 \\
 0 0 &0 0 \underline{1} 0& 0 0 0 0 & 0 1 0 0 0 \\
 0 0 &0 0 0 \underline{1}& 0 0 0 0 & 1 1 0 0 0
\end{tabular}
\right ).
$$
(Note that, in this case, $\tau$ has no relation to the orbit decomposition.)
Then, we permute the corresponding matrix $[{-A}^T| I] $ by $\tau^{-1}$ to obtain the generator matrix
$G_2$ of $\mathcal{C}_2$.
$$G_2=\left(
\begin{tabular}{c|c|c}
 1 1 1 0 0 0 & 0 0 0 0 0 0 & 0 0 0\\
 1 0 0 1 0 0 & 0 0 0 0 0 0 & 0 0 0\\
 0 1 0 0 1 0 & 0 0 0 0 0 0 & 0 0 0\\
 1 1 0 0 0 1 & 0 0 0 0 0 0 & 0 0 0\\
 1 1 0 0 0 0 & 1 1 0 1 1 0 & 0 0 0\\
 0 1 0 0 0 0 & 1 0 1 1 0 1 & 0 0 0\\
 1 0 0 0 0 0 & 0 0 0 0 0 0 & 1 0 0\\
 0 1 0 0 0 0 & 0 0 0 0 0 0 & 0 1 0\\
 1 1 0 0 0 0 & 0 0 0 0 0 0 & 0 0 1
\end{tabular}
\right)$$
By using Buchberger's algorithm, we can compute the reduced POT
Gr\"obner basis $\left\{g_1,g_2,g_3\right\}$ of generator matrix $G_2$:
\begin{center}
$
\left[\begin{array}{c}
  g_1 \\
  g_2 \\
  g_3 \end{array} \right] =
\left[ \begin{array}{ccc}
  1, & 0, & 1 \\
  0, &1+t+t^3+t^4, & 1+t \\
  0, & 0, & 1+t+t^2 \end{array} \right]. \nonumber
$
\end{center}
\hfill$\Box$

Although this example is binary, our algorithm can be applied to all parity check matrix $H$ of $\mathbb{F}_q$-entries.
We consider in section \ref{algorithms} the computational complexity of our algorithm to obtain $\mathcal{G}$ from $H$.

\section{Transpose formula for POT Gr\"obner basis}
\label{generalized}
In this section, we propose another algorithm to compute the Gr\"obner basis from parity check matrix.
This novel algorithm uses transpose formula \eqref{formula} that is given at Theorem \ref{theorem 2}.
Although Theorem 1 is not necessary for our computation except a scalar product \eqref{inner} and Corollary \ref{Corollary}, we describe it for completeness;
Theorem 1 provides the orthogonal property with respect to the scalar product for arbitrary Gr\"obner bases of GQC codes.

Firstly, we define a circulant $l \times l$ matrix as a square $l \times l$ matrix such that each row is constructed from the previous row by a single right cyclic shift.
Then, we can represent the circulant $l \times l$  matrix
$$
A  = \left( \begin{array}{cccc}
a_0&a_1&\cdots&a_{l-1} \\
a_{l-1}&a_0&\cdots&a_{l-2} \\
\vdots & \ddots & \ddots & \vdots\\
a_1&\cdots&a_{l-1}&a_{0}
\end{array}\right)
$$
as a polynomial
$a(t) = a_0  + a_1 t + \cdots + a_{l - 1} t^{l - 1} $ in module $\mathbb{F}_q[t]/ \left ( {t^l-1} \right)$.
\begin{proposition}\label{proposition}
Let $a(t)$ and $b(t)$ represent the corresponding polynomials of circulant matrices $A$ and $B$ of size $l\times l$, respectively.\\
(i) Transpose of $A$ is a circulant matrix corresponding to polynomial $\widehat{a}(t):= a_0  + a_{l - 1} t + \cdots + a_1 t^{l - 1}$ in module $\mathbb{F}_q[t]/ \left ( {t^l-1} \right)$.\\
(ii) Matrix product $AB$ equals $BA$ and corresponds to polynomial $a(t)b(t)$.
In particular, we have $AB=0$ if and only if $a(t)b(t)\equiv0$ mod $\left(t^l-1 \right)$.\\
(iii) If $\left[{A_1 \cdots A_m}\right]*
\left[ \begin{array}{c}
                B_1^T \\
                \vdots  \\
                B_m^T
           \end{array} \right]=0$ holds, where $A_i$ and $B_i$ are circulant $l\times l$ matrices, then we have the corresponding polynomial $\sum\limits_{i = 1}^m a_i(t)\widehat{b}_i(t)\equiv0$ mod $\left (t^l-1 \right)$.\\
(iv) The product $\left[ \begin{array}{c}
               A  \\
              \vdots  \\
               A
           \end{array} \right]*
\left[ B^T \cdots B^T \right]$ equals a circulant matrix of size $ml \times ml$ and corresponds to polynomial
$a(t)\widehat{b}(t)\sum\limits_{i = 0}^{m-1} {t^{il}}$ mod $\left (t^{ml}-1 \right )$.
\end{proposition}
\emph{Proof:} Proposition 1.(i)--(iii) are easy to prove and we refer to \cite{Lally}.
Proposition 1.(iv) can be proved by executing matrix multiplication
\begin{equation}\label{executing}
 \left[ \begin{array}{c}
    A \\
  \vdots  \\
    A
 \end{array} \right]
 \left[ \begin{array}{c}
  B^T \cdots B^T
 \end{array} \right]
 =\left[ \begin{array}{ccc}
    AB^T & \cdots & AB^T \\
   \vdots & \cdots & \vdots   \\
    AB^T & \cdots & AB^T
 \end{array} \right].
\end{equation}
From (i) and (ii), we see that the matrix \eqref{executing} is a circulant matrix of size $ml \times ml$, and moreover, the circulant matrix ${AB^T}$ can correspond to polynomial $a(t)\widehat{b}(t)$ in module $\mathbb{F}_q[t]/ \left ( {t^{l}-1} \right)$.
Therefore, matrix \eqref{executing} can be represented by the following polynomial
$$
 a(t)\widehat{b}(t)\left( {1+t^l+\cdots+t^{l(m-1)} }\right )=
 a(t)\widehat{b}(t)\sum\limits_{i = 0}^{m-1} { t^{il} }
$$
in module $\mathbb{F}_q[t]/ \left ({t^{ml}-1} \right)$.
\hfill $\Box$
\vspace{1em}

Next, we define a scalar product of polynomial vectors $u = (u_1,\cdots, u_m),\\v = (v_1,\cdots, v_m)\in M$ as
\begin{equation}\label{inner}
\left\langle{u,v}\right\rangle :=
 \sum\limits_{i = 1}^m{u_i (t)\widehat{v}_i(t)\sum\limits_{k=0}^{l/l_i-1}{t^{kl_i}}} \; \bmod \left({t^{l}-1}\right),
\end{equation}
where $l$ is the least common multiple (lcm) of $l_i$'s that correspond to non-zero $u_i$ and $v_i$.
We denote $[u]$ as the matrix representation of polynomial vector $u=(u_{1},\cdots,u_{m})\in M$ by shifting locally cyclically $l$ times.
Since $l_i$ divides $l$ and $t^{l_i}$ is regarded as 1, we can represent matrix $[u]$ by non-zero circulant matrices $[u_{i}]$ and zero matrices.
For example, assume $u=(1+t,0,1+t^2),\,v=(1+t+t^3, 1+t, 1+t)\in M$, where $l_1=5,\,l_2=4,\,l_3=3,\,q=2$.
Since $l = \mathrm{lcm}(l_1,l_3) = 15$, the matrix representation of
polynomial vector $u$ agrees with
$$
\left[
     \begin{tabular}{c}
        1  1  0  0  0 \\
        0  1  1  0  0 \\
        0  0  1  1  0 \\
        0  0  0  1  1 \\
        1  0  0  0  1 \\
        \hline
        1  1  0  0  0 \\
        0  1  1  0  0 \\
        0  0  1  1  0 \\
        0  0  0  1  1 \\
        1  0  0  0  1 \\
        \hline
        1  1  0  0  0 \\
        0  1  1  0  0 \\
        0  0  1  1  0 \\
        0  0  0  1  1 \\
        1  0  0  0  1
      \end{tabular}
      \right]
         \left[
          \begin{tabular}{c}
             0 0  0  0 \\
             0 0  0  0 \\
             0 0  0  0 \\
             0 0  0  0 \\
             0 0  0  0 \\
             0 0  0  0 \\
             0 0  0  0 \\
             0 0  0  0 \\
             0 0  0  0 \\
             0 0  0  0 \\
             0 0  0  0 \\
             0 0  0  0 \\
             0 0  0  0 \\
             0 0  0  0 \\
             0 0  0  0
          \end{tabular}
         \right]
         \left[
          \begin{tabular}{c}
             1  0  1 \\
             1  1  0 \\
             0  1  1 \\
             \hline
             1  0  1 \\
             1  1  0 \\
             0  1  1 \\
             \hline
             1  0  1 \\
             1  1  0 \\
             0  1  1 \\
             \hline
             1  0  1 \\
             1  1  0 \\
             0  1  1 \\
             \hline
             1  0  1 \\
             1  1  0 \\
             0  1  1
          \end{tabular}
         \right].
$$
We see that matrix $[u]$ can be decomposed into two non-zero circulant matrices
$$\left(
   \begin{array}{ccccc}
      1 & 1 & 0 & 0 & 0 \\
      0 & 1 & 1 & 0 & 0 \\
      0 & 0 & 1 & 1 & 0 \\
      0 & 0 & 0 & 1 & 1 \\
      1 & 0 & 0 & 0 & 1
   \end{array}
\right)\;\mbox{and}\;
 \left(
     \begin{array}{ccc}
       1 & 0 & 1 \\
       1 & 1 & 0 \\
       0 & 1 & 1
     \end{array}
  \right ) ,
$$
which correspond to polynomials $(1+t)$ mod $\left(t^5-1\right)$, and $\left(1+t^2\right)$ mod $\left(t^3-1\right)$, respectively.
The scalar product $\left \langle u,v \right \rangle$ agrees with
\begin{gather*}
 {(1+t)(1+t^2+t^4)\sum\limits_{k=0}^{2}{t^{5k}} + (1+t^2)(1+t^2)\sum\limits_{k=0}^{4}{t^{3k}}}\\
\equiv(1+t^{2}+t^{8}+t^{10}+t^{11}+t^{14})
\mod\left({t^{15}-1}\right).
\end{gather*}
With these preparations, we can obtain the following orthogonality between a Gr\"obner basis of a GQC code and that of its dual.

\begin{theorem}\label{theorem 1}
Let $ \mathcal{G} = \left \{ {g_1,\cdots,g_m } \right\} $ and
$\mathcal{H}=\left \{ {h_1,\cdots,h_m} \right \}$ be a Gr\"obner basis of a GQC code $\mathcal{C}$ and that of $\mathcal{C}^\bot$ with
respect to any ordering, respectively. Then, we have
$\left\langle{g_i,h_j}\right\rangle\equiv\left\langle{h_j,g_i}\right\rangle\equiv0$ mod $\left({t^{l}-1}\right)$ for all $1 \le i,j \le m$.
\end{theorem}
\emph{Proof:}
$\left\langle{h_j,g_i}\right\rangle\equiv0$ follows from $\left\langle{g_i,h_j}\right\rangle\equiv0$ easily.
Shifting the component $g_i=(g_{i1}, \cdots, g_{im})$ locally cyclically $l$ times,
we obtain
$ \left\{ {g_i, tg_i, \cdots , t^{l-1}g_i} \right\}$ that correspond to $l$ polynomial vectors as follows:
\begin{equation}\label{shift}
\left( \begin{array}{cccc}
  g_{i1}  & g_{i2}  &  \cdots & g_{im}  \\
  tg_{i1} & tg_{i2} &  \cdots & tg_{im}  \\
  \vdots  &  \vdots &  \cdots & \vdots  \\
  t^{l-1 } g_{i1} & t^{l-1 } g_{i2} &  \cdots & t^{l-1} g_{im}
 \end{array} \right).
\end{equation}
Let $[g_i]$ denote the matrix corresponding to $l$ vectors \eqref{shift}.
We can represent $[g_i]$ by circulant matrices
$\left [ g_{ij} \right ]$ as follows:
\begin{equation}
\left[ {g_{i} } \right] =
\left ( {
     \begin{array}{l}
       \left[ {g_{i1} } \right] \\
            \;\; \vdots  \\
       \left[ {g_{i1} } \right] \\
      \end{array}
          \begin{array}{l}
             \left[ {g_{i2} } \right] \\
                \;\; \vdots  \\
             \left[ {g_{i2} } \right] \\
          \end{array}
          \cdots
      \begin{array}{l}
             \left[ {g_{im} } \right] \\
             \;\; \vdots  \\
             \left[ {g_{im} } \right] \\
             \end{array}
      } \right),
\end{equation}
where
$\left( \begin{array}{l}
 \left[ {g_{ik} } \right] \\
 \;\; \vdots  \\
 \left[ {g_{ik} } \right] \\
 \end{array} \right)$
is the $l \times l_k $ matrix made from non-zero matrix $\left[{g_{ik}}\right]$ or only from zeros.
Since $t^{\delta}g_i \in \overline{\mathcal{C}}$ for all $\delta$, every rows of $\left[ {g_{i} } \right]$ are codewords in $\mathcal{C}$.
Similarly, the corresponding matrix representation of $h_j $ agrees with
$$ \left[ h_j \right] =
\left( {
 \begin{array}{l}
 \left[ {h_{j1} } \right] \\
 \;\; \vdots  \\
 \left[ {h_{j1} } \right]
 \end{array}
 \begin{array}{l}
 \left[ {h_{j2} } \right] \\
 \;\; \vdots  \\
 \left[ {h_{j2} } \right]
 \end{array}
 \cdots
 \begin{array}{l}
 \left[ {h_{jm} } \right] \\
 \;\; \vdots  \\
 \left[ {h_{jm} } \right]
 \end{array} }
 \right)$$
and every rows of $\left[ {h_{j} } \right]$ are codewords in $\mathcal{C}^{\bot}$.
The relation $c * \left( { c^{\bot} } \right )^T = 0$,
where $c \in\mathcal{C}$ and $ c^{\bot} \in\mathcal{C}^{\bot}$, corresponds to
$
  \left[ {g_{i} } \right]\left[ {h_{j} } \right]^T \!= 0
$ for all $i,j$.
Therefore, we have the following equivalent equation
\begin{align}
&\left( \begin{array}{l}
    [g_{i1}] \\
     \;\; \vdots \\
    {[g_{i1}]}
 \end{array} \right)
 \left( \begin{array}{l}
 [h_{j1}] \\
 \;\; \vdots  \\
 {[h_{j1}]}
 \end{array} \right)^T
 +
 \left( \begin{array}{l}
    [g_{i2}] \\
     \;\; \vdots \\
    {[g_{i2}]}
 \end{array} \right)
 \left( \begin{array}{l}
 [h_{j2}] \\
 \;\; \vdots  \\
 {[h_{j2}]}
 \end{array} \right)^T\nonumber\\
& + \cdots  + \left( \begin{array}{l}
 [g_{im}] \\
 \;\; \vdots  \\
 {[g_{im}]} \\
 \end{array} \right)
 \left( \begin{array}{l}
 [h_{jm}] \\
 \;\; \vdots  \\
 {[h_{jm}]}
 \end{array} \right)^T = 0.
\label{orthogonal}
\end{align}
If $ g_{ik}=0 $ or $h_{ik}=0$, then the $k$-th term of \eqref{orthogonal} is zero matrix of size $l \times l$.
Otherwise, by Proposition \ref{proposition}.(iv), the $k$-th term of \eqref{orthogonal} is a circulant $l \times l$ matrix and the corresponding polynomial agrees with
$$
g_{ik}(t)\widehat{h}_{jk}(t)\left(t^l-1\right)/\left(t^{l_k}-1\right)
\mod\;\left(t^{l}-1\right).
$$
Therefore, the corresponding polynomial of \eqref{orthogonal} is obtained as follows:
\begin{align*}
&g_{i1}(t)\widehat{h}_{j1}(t)\sum\limits_{\delta=0}^{l/l_1-1}{t^{\delta l_1}}
 + g_{i2}(t)\widehat{h}_{j2}(t)\sum\limits_{\delta=0}^{l/l_2-1}{t^{\delta l_2}}\\
&+\cdots+ g_{im}(t)\widehat{h}_{jm}(t)\sum\limits_{\delta=0}^{l/l_m-1}{t^{\delta l_m}}
  = \left\langle {g_i ,h_j } \right\rangle \!\equiv\! 0
\end{align*}
modulo $\left(t^{l}-1\right)$, which leads the theorem.
\hfill$\Box$

By using Theorem \ref{theorem 1}, we can compute the Gr\"obner basis $\mathcal{G}$ from $\mathcal{H}$.
However, the computation is not straightforward because of the ambiguity ``$\mathrm{mod}\,\left(t^{l}-1\right)$.''
Little \emph{et al.} \cite{Little} obtained strict equalities for POT and rPOT diagonal components $ \left \langle g_i,h_i\right \rangle$, which we applied to finite geometry codes in \cite{Van}.
Now, we remove all modulo conditions.
For later use, we derive a corollary from the argument at \eqref{orthogonal}.

\begin{corollary}\label{Corollary}
Let $\mathcal{H}=\left \{ {h_1,h_2,\cdots,h_m} \right \}$ be a Gr\"obner basis of $\overline{\mathcal{C}^\bot}$, and $u \in M$ a polynomial vector. Then, it holds that $ \left \langle {h_i, u } \right \rangle\equiv0$ for all $1 \le i \le m$ if and only if $u$ corresponds to a codeword in $\mathcal{C}$.
\hfill$\Box$
\end{corollary}

In the case of cyclic codes, if we know the generator polynomial $h(t)$ of the dual code $\mathcal{C}^\bot$ and $a(t)h(t)=t^n-1$, then that of $\mathcal{C}$ is the reciprocal polynomial $t^{\deg a}a(t^{-1})$ of $a(t)$, which agrees with $t^{\deg a\,}\widehat{a}(t)\mod\left(t^n-1\right)$.
We generalize this relation to GQC codes.
Assume that $\mathcal{H}$ is rPOT Gr\"obner basis of an $m$-orbit GQC code $\mathcal{C}^\bot$.
Since $\mathcal{H}$ is a basis of $\overline{\mathcal{C}^\bot}$ (as described at \eqref{generation}), there exists $m\times m$ polynomial matrix $A=(a_{ij})$ satisfying
$$
A
\left[
   \begin{array}{c}
h_1 \\
h_2 \\
\vdots \\
h_m
   \end{array}
\right]
=
\left[
   \begin{array}{cccc}
     t^{l_1}-1&  0 & \cdots & 0  \\
       0     & t^{l_2}-1 & \ddots &\vdots \\
     \vdots & \ddots  & \ddots  & 0 \\
       0     & \cdots    &  0 &t^{l_m}-1
   \end{array}
\right].
$$
It is easy to observe that $A=(a_{ij})$ is a lower triangular matrix similar to $(h_{ij})$, namely, $a_{ij}=0$ if $i<j$.
If $\mathcal{C}^\bot$ (or $\mathcal{C}$) is a QC code, then we have $A(h_{ij})=(h_{ij})A$ as noticed in \cite{Lally}, but in general not commutative.
We can calculate $a_{ij}$ recursively as follows:
\begin{equation}\label{successive}
    a_{ij} := \left\{  \begin{array}{ccl}
                         0 & \mathrm{if} &j>i, \vspace{1mm}\\
\dfrac{t^{l_i}-1}{h_{ii}}  & \mathrm{if} &j=i,  \vspace{2mm}\\
\displaystyle\frac{-1}{h_{jj}}\sum_{\delta = j + 1}^i {a_{i \delta} h_{\delta j} } & \mathrm{if} &j<i.  \\
                       \end{array}
            \right.
\end{equation}

It is important fact that, if $\mathcal{H}$ is the reduced rPOT Gr\"obner basis, then $A=(a_{ij})$ has the similar property, that is, $\deg a_{ij} < \deg a_{ii}$ for all $i > j$.
Now we prove this by induction on $j$.
The first step $\deg a_{i,i-1} < \deg a_{ii}$ follows from $a_{i,i-1}h_{i-1,i-1} + a_{ii}h_{i,i-1} = 0$ from \eqref{successive}.
Suppose induction hypothesis $\deg a_{i\delta} < \deg a_{ii}$ for $j+1\le\delta<i$.
From \eqref{successive}, we obtain
\begin{align*}
\deg a_{ij}
&=\deg\left(
\sum_{\delta = j + 1}^i a_{i \delta} h_{\delta j}
\right)-\deg h_{jj}\\
&\le\max_{j<\delta\le i}
\left\{\deg a_{i\delta}+\deg h_{\delta j}-\deg h_{jj}\right\}\\
&<\deg a_{ii},
\end{align*}
which proves the fact.

From now on, we assume that $\mathcal{H}$ is the reduced rPOT Gr\"obner basis.
We define transpose polynomial matrix of $A$ by
\begin{equation}\label{transpose}
\left[ \begin{array}{cccc}
\widehat{a}_{11} & \widehat{a}_{21} & \cdots & \widehat{a}_{m1} \\
0 & \widehat{a}_{22} & \cdots & \widehat{a}_{m2}  \\
\vdots & \ddots & \ddots &\vdots          \\
0& \cdots& 0& \widehat{a}_{mm}
                     \end{array}
\right]=:
\left[
   \begin{array}{c}
b_1 \\
b_2 \\
\vdots \\
b_m
   \end{array}
\right],
\end{equation}
where $\widehat{a}_{ij}$ is calculated in $\mathbb{F}_q[t]/ \left ( {t^{l_i}-1} \right)$, and not in $\mathbb{F}_q[t]/ \left ( {t^{l_j}-1} \right)$.
Since $a_{ij}$ is the $j$-th component of a polynomial row vector, it might seem natural to calculate $\widehat{a}_{ij}$ in $\mathbb{F}_q[t]/ \left ({t^{l_j}-1} \right)$.
Nevertheless, we consider $\widehat{a}_{ij}$ modulo $\left ({t^{l_i}-1}\right)$, which is justified by $\deg a_{ij}<\deg a_{ii}\le l_i$ and is a characteristic of GQC codes that is disappeared in the case of QC codes.

The latter half of the following theorem provides the objective formula of POT Gr\"obner basis for GQC codes.
\begin{theorem}\label{theorem 2}
Polynomial vectors \eqref{transpose} satisfy
\begin{equation}\label{main}
\left \langle {h_i,b_j} \right \rangle=\left \{
\begin{array}{cl}
 t^{l_i}-1 & \;1 \le i = j \le m,\\
  0 & \;1 \le i \ne j \le m. \\
\end{array}
\right.
\end{equation}
Moreover, $\mathcal{G}=\{g_1,\cdots,g_m\}$, where
\begin{equation}\label{formula}
g_{ij} := t^{\deg a_{ii}}b_{ij} \mod\left(t^{l_i}-1\right),
\end{equation}
determines a POT Gr\"obner basis of GQC code $\mathcal{C}$ (usually not reduced).
\end{theorem}

This formula \eqref{formula} generalizes that of cyclic codes and that of QC codes by Lally--Fitzpatrick \cite{Lally} to the case of GQC codes.
In \cite{Lally}, their formula is proved by the fundamental row operation of a polynomial matrix; this proof cannot be applied to our case because of the complication to different orbit lengths.
We first show \eqref{main} directly from \eqref{successive}, then we conclude by degree argument.
\vspace{1em}

\noindent
\emph{Proof of Theorem \ref{theorem 2}:}
From the definition \eqref{inner}, it obviously holds that\\
$\left\langle h_i,b_j \right\rangle=0$ for $i<j$, and then we concentrate on the case of $j\le i$.
Consider two polynomial vectors
$h_{i} = \left (h_{i1},\cdots,h_{ii}, 0, \cdots , 0 \right)\in\mathcal{H}$ and $b_{j} = \left (0,\cdots,0, \widehat{a}_{jj},\cdots,\widehat{a}_{mj}\right)$, where $1 \le j \le i \le m$.
From \eqref{successive} and $\widehat{\widehat{a}}=a$, it is trivial that $\left \langle {h_k,b_k} \right \rangle =t^{l_k}-1$ for all $1 \le k \le m$.
Thus, we may prove only $\left \langle {h_i,b_j} \right \rangle = 0$ for all $j<i$ by induction on $i-j$.
We denote ${\beta_{ij}}:=\mathrm{lcm}(l_j,\cdots,l_i)$, then $\left \langle{ h_i, b_j}\right \rangle$ is computed as follows:
\begin{equation}\label{production}
\left \langle {h_i,b_j} \right \rangle = \left(t^{\beta_{ij}} -1\right)
\sum_{j \le k \le i}\frac{h_{ik}a_{kj}}{t^{l_k}-1}.
\end{equation}
If $i-j=1$, we have
\begin{align*}
\left \langle {h_i,b_j} \right \rangle &=
\left(t^{\beta_{ij}}-1\right)\left[ \frac{ h_{ij}a_{jj} }{t^{l_j}-1} +
\frac{ h_{ii}a_{ij} }{t^{l_i}-1} \right]
\\
&= \frac{t^{\beta_{ij}}-1}{h_{jj}a_{ii}}( h_{ij}a_{ii} + h_{jj}a_{ij})
= 0.
\end{align*}
If $i-j=2$, we have
\begin{align*}
\left \langle {h_i,b_j} \right \rangle &=
\left(t^{\beta_{ij}} - 1\right)\left[ {\frac{h_{ij}a_{jj}}{t^{l_j}-1}
+ \frac{h_{i,j+1}a_{j+1,j}}{t^{l_{j+1}}-1}
+ \frac{h_{ii}a_{ij}}{t^{l_i}-1} } \right] \nonumber \\
&= \left(t^{\beta_{ij}}-1\right)\left[{\frac{h_{ij}}{h_{jj}}
+\frac{h_{i,j+1}a_{j+1,j}}{h_{j+1,j+1}a_{j+1,j+1}} + \frac{a_{ij}}{a_{ii}} }\right].
\end{align*}
From \eqref{successive}, we obtain
$$
  h_{i,j+1} = \frac{-a_{i,j+1}h_{j+1,j+1}}{a_{ii}}, \quad  a_{j+1,j} = \frac{-a_{j+1,j+1}h_{j+1,j}}{h_{jj}}.
$$
Therefore, $\left\langle h_i,b_j\right\rangle$ corresponds to
$$
\frac{t^{\beta_{ij}}-1}{h_{jj}a_{ii}} \left({a_{ij}h_{jj} + a_{i,j+1}h_{j+1,j} + a_{ii}h_{ij} }\right) = 0.$$
Suppose induction hypothesis $\left \langle{ h_\theta, b_\delta }\right \rangle=0$ for all $\theta<\delta$ with $\delta-\theta<i-j$.
From \eqref{successive}, we receive, for all $j+1 \le k \le i-1$, the following equations
$$
h_{ik} = \frac{-1}{a_{ii}}\sum\limits_{\delta = k}^{i - 1} {a_{i\delta}h_{\delta k}} , \quad
a_{kj} = \frac{-1}{h_{jj}}\sum\limits_{\theta = j + 1}^k {a_{k\theta} h_{\theta j}}.
$$
Consider the following partial summation of \eqref{production}:
\begin{align}
a_{ii}h_{jj}\sum_{k=j+1}^{i-1}\frac{h_{ik}a_{kj}}{t^{l_k}-1}
  &=\sum_{k=j+1}^{i - 1}\frac{1}{t^{l_k }  - 1}
  \left( \sum_{\delta=k}^{i-1}
a_{i \delta} h_{\delta k} \right)
  \left( \sum_{\theta=j+1}^{k}
a_{k\theta} h_{\theta j}  \right)
\nonumber\\
&=
\sum_{j+1\le k\le i - 1} \sum_{\scriptstyle k \le \delta  \le i - 1 \hfill \atop
  \scriptstyle j+1 \le \theta  \le k \hfill} \frac{1}{ t^{l_k }- 1 } a_{i\delta} h_{\delta k} a_{k\theta } h_{\theta j}
\nonumber\\
\label{right-hand}
&=
\sum\limits_{\scriptstyle j+1 \le \theta\le\delta  \le i - 1}
a_{i\delta} h_{\theta j} \sum\limits_{\theta  \le k \le \delta }
\frac{1}{
 t^{l_k } - 1 }h_{\delta k} a_{k\theta }.
\end{align}
For all $j+1 \le \theta < \delta \le i-1$, we have $\left \langle{ h_\theta, b_\delta }\right \rangle=0$ by induction hypothesis.
Therefore, from \eqref{successive}, the double summation \eqref{right-hand} is equal to
$$
\sum_{\theta=j+1}^{i - 1} a_{i\theta} h_{\theta j} \frac{h_{\theta\theta} a_{\theta\theta} }{t^{l_\theta }  - 1 }
= \sum_{\theta=j+1}^{i - 1} a_{i\theta} h_{\theta j}
=-a_{ij} h_{jj}-a_{ii} h_{ij}.
$$
From this result, $\left\langle h_i,b_j \right\rangle$ is equal to
$$
\left(t^{\beta_{ij}} -1\right)\left[
\frac{h_{ij}a_{jj}}{t^{l_j}-1}
-\frac{a_{ij}}{a_{ii}}-\frac{h_{ij}}{h_{jj}}+
\frac{h_{ii}a_{ij}}{t^{l_i}-1}\right]\\
=0
$$
Therefore, we have $\left \langle {h_i,b_j} \right \rangle = 0$ for all $1 \le i \ne j \le m$.

The rest of the proof is to show that \eqref{formula} determines a POT Gr\"obner basis in the meaning of Definition \ref{Grobner}.
From Corollary \ref{Corollary}, we have $b_i\in\mathcal{C}$, then $g_i\in\mathcal{C}$.
Thus, we may prove only that $g_{ii}$ has the minimum degree among the vectors of the form $(0,\cdots,0,c_i(t),\cdots,c_m(t))\in\overline{\mathcal{C}}$ with $c_i(t)\not=0$.
We notice that $\mathcal{C}_i:=\left\{c_i\,|\,(0,\cdots,0,c_i,\cdots,c_m)\in\mathcal{C}\right\}$ defines a cyclic code.
Since the generator polynomial of the dual code $\mathcal{C}_i^\bot$ is $h_{ii}$, that of $\mathcal{C}_i$ is $g_{ii}$, then $g_{ii}$ has the minimum degree.
\hfill $\Box$

\begin{figure*}[t!]
\begin{center}
  \resizebox{13.5cm}{!}{\includegraphics{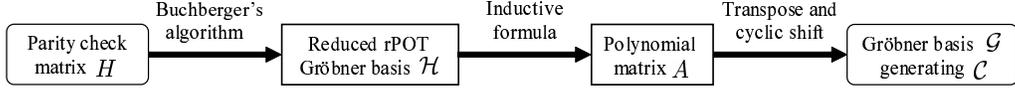}}
\caption{Outline of computing Gr\"obner basis by transpose algorithm from parity check matrix.\label{Lally-Fitzpatrick}}
\end{center}
\end{figure*}

By Theorem \ref{theorem 2}, we obtain the second algorithm for computing Gr\"obner basis $\mathcal{G}$ of $m$-orbit GQC code $\mathcal{C}$ from the parity check matrix as follows.
The flow of this algorithm is presented in Figure \ref{Lally-Fitzpatrick}.

\begin{Lally}\hfill
{\rm
\\
\textbf{Input:} Parity check matrix $H$ of a GQC code $\mathcal{C}$.
\\
\textbf{Output:} POT Gr\"obner basis $\mathcal{G}$ of $\overline{\mathcal{C}}$.
\\
\textbf{Step 1.} Compute the reduced rPOT Gr\"obner basis $\mathcal{H}$ by Buchberger's algorithm
from matrix $H$.
\\
\textbf{Step 2.} Calculate $A=(a_{ij})$ by \eqref{successive}.\\
\textbf{Step 3.} Obtain
$\mathcal{G} = \left\{g_1,\cdots,g_m\right\}$, where
$$
g_i=(g_{ij})_{1 \le j \le m},\quad
    g_{ij} := \left\{  \begin{array}{cl}
0 & \mathrm{if}\;\,i > j,   \\
t^{\deg a_{ii}}\widehat{a}_{ji}  & \mathrm{if}\;\,i \le j. \qquad\Box
                       \end{array}
            \right.
$$
}
\end{Lally}

\begin{remark}
{\rm
We can construct the generator matrix $G$ of GQC code $\mathcal{C}$ from its reduced POT Gr\"obner basis $\left \{g_1,\cdots,g_m \right \}$ as follows:
}
\end{remark}
\begin{equation}\label{generator matrix}
G=\left( \begin{array}{llll}
          g_{11}  & g_{12}  & \cdots  & g_{1m}  \\
          tg_{11} & tg_{12} & \cdots  & tg_{1m} \\
          \vdots  & \vdots  & \vdots  & \vdots  \\
          t^{x_1}g_{11} & t^{x_1}g_{12} & \cdots  & t^{x_1}g_{1m} \\
          0       & g_{22}  & \cdots  & g_{2m}    \\
          0       & tg_{22}  & \cdots  & tg_{2m}  \\
          \vdots  & \vdots  & \vdots  & \vdots    \\
          0       & t^{x_2}g_{22} & \cdots  & t^{x_2}g_{2m} \\
          \vdots  & \vdots  & \ddots  & \vdots  \\
          0       & 0  & \cdots  & g_{mm}    \\
          0       & 0  & \cdots  & tg_{mm}  \\
          \vdots  & \vdots  & \vdots  & \vdots    \\
          0       & 0 & \cdots  & t^{x_m}g_{mm}
     \end{array}
 \right),
\end{equation}
where $g_{i}= \left(0,\cdots,0,g_{ii},\cdots, g_{im}\right)$ and
$x_i:=l_i-\deg g_{ii}-1$ for all $1 \le i \le m$.
Since the diagonal components $g_{ii}$ all lie in different position, the rows of this matrix are linearly independent.
Moreover, the total number of rows equals $\sum\limits_{i = 1}^m {\left( {l_i - \deg g_{ii} } \right)} = k$.
Therefore, the matrix \eqref{generator matrix} provides the generator matrix of GQC code $\mathcal{C}$, which generalizes the representation for quasi-cyclic codes in \cite{Lally}.
\hfill$\Box$

\begin{example}
{\rm
We demonstrate the transpose algorithm.
Let $\mathcal{C}_3$ be a binary GQC code with $l_1 = 6,\,l_2 = 6,\,l_3 = 4$ defined by
$$
H_3:=\left(
\begin{array}{c|c|c}
 0 \;\; 1 \;\; 1 \;\; 0 \;\; 1 \;\; 1&  1 \;\; 0 \;\; 1 \;\; 0 \;\; 0 \;\; 0&  0 \;\; 0 \;\; 0 \;\; 0 \\
 1 \;\; 0 \;\; 1 \;\; 1 \;\; 0 \;\; 1&  0 \;\; 1 \;\; 0 \;\; 1 \;\; 0 \;\; 0&  0 \;\; 0 \;\; 0 \;\; 0 \\
 1 \;\; 1 \;\; 0 \;\; 1 \;\; 1 \;\; 0&  0 \;\; 0 \;\; 1 \;\; 0 \;\; 1 \;\; 0&  0 \;\; 0 \;\; 0 \;\; 0 \\
 0 \;\; 1 \;\; 1 \;\; 0 \;\; 1 \;\; 1&  0 \;\; 0 \;\; 0 \;\; 1 \;\; 0 \;\; 1&  0 \;\; 0 \;\; 0 \;\;0 \\
 \hline
 1 \;\; 1 \;\; 0 \;\; 1 \;\; 1 \;\; 0&  1 \;\; 0 \;\; 0 \;\; 0 \;\; 0 \;\; 0&  1 \;\; 0 \;\; 0 \;\; 0 \\
 0 \;\; 1 \;\; 1 \;\; 0 \;\; 1 \;\; 1&  0 \;\; 1 \;\; 0 \;\; 0 \;\; 0 \;\; 0&  0 \;\; 1 \;\; 0 \;\; 0 \\
 1 \;\; 0 \;\; 1 \;\; 1 \;\; 0 \;\; 1&  0 \;\; 0 \;\; 1 \;\; 0 \;\; 0 \;\; 0&  0 \;\; 0 \;\; 1 \;\; 0 \\
 1 \;\; 1 \;\; 0 \;\; 1 \;\; 1 \;\; 0&  0 \;\; 0 \;\; 0 \;\; 1 \;\; 0 \;\; 0&  0 \;\; 0 \;\; 0 \;\; 1
\end{array}
\right).
$$
We calculate the reduced rPOT Gr\"obner basis $\mathcal{H}_3=\left \{ {h_1,h_2,h_3} \right \}$ of dual code $\mathcal{C}_3^\bot$ by Buchberger's algorithm:
$$
\left[ \begin{array}{c}
h_1\\h_2\\h_3
\end{array}\right]
=
\left[ \begin{array}{ccc}
 1 + t^6,&0,& 0 \\
 t+t^2+t^4+t^5,& 1+t^2,&0 \\
 1+t+t^3+t^4,&1, & 1
 \end{array} \right].
$$
There exists a polynomial matrix $A=(a_{ij})$ satisfying $A[h_i] = 0$.
From \eqref{successive}, we can calculate $A$ inductively:
$$
A =
\left[ \begin{array}{ccc}
    1, & 0, & 0  \\
    t+t^2+t^3, & 1+t^2+t^4, & 0\\
    1+t^2, & 1+t^2, &  1+t^4
\end{array} \right].
$$
The transpose polynomial matrix of $A$ turns into
$$
\left[ \begin{array}{c}
    \widehat{a}_{11}, \widehat{a}_{21} , \widehat{a}_{31}  \\
    \;0, \;\:\widehat{a}_{22}, \:\widehat{a}_{32}\\
    \;0, \;\;\:0, \;\;\:\widehat{a}_{33}
\end{array} \right]
=
\left[ \begin{array}{ccc}
    1, & t^3+t^4+t^5, & 1+t^2  \\
    0, & 1+t^2+t^4, & 1+t^2\\
    0, & 0, & 1+t^4
\end{array} \right].
$$
According to Theorem \ref{theorem 2}, a POT Gr\"obner basis of GQC code $\mathcal{C}_3$ can be computed by $g_{ij} := t^{\deg a_{ii}\,}\widehat{a}_{ji} \mod \left(t^{l_i}-1\right)$.
After reduction, we obtain the reduced POT Gr\"obner basis $\mathcal{G}_3=\left \{ g_1,g_2,g_3 \right \}$:
\begin{equation}\label{result}
\left[ \begin{array}{c}
g_1\\g_2\\g_3
\end{array}\right]
=
\left[ \begin{array}{ccc}
 1,& 1+t+t^2,& t+t^3 \\
 0,& 1+t^2+t^4, & 1+t^2 \\
 0, & 0, & 1+t^4
 \end{array} \right].
\end{equation}
To check the correctness of \eqref{result}, we calculate the generator matrix $G_3$ of $\mathcal{C}_3$ by \eqref{generator matrix}:
$$
G_3=\left(
\begin{array}{c|c|c}
 1 \;\; 0 \;\; 0 \;\; 0 \;\; 0\;\; 0 & 1\;\; 1\;\; 1\;\; 0\;\; 0\;\; 0 & 0\;\; 1\;\; 0\;\; 1 \\
 0 \;\;1 \;\;0 \;\;0 \;\;0 \;\;0 & 0\;\; 1 \;\;1 \;\;1 \;\;0 \;\;0 & 1\;\; 0 \;\;1 \;\;0 \\
 0 \;\;0 \;\;1 \;\;0\;\; 0\;\; 0 & 0 \;\;0 \;\;1\;\; 1\;\; 1\;\; 0 & 0\;\; 1\;\; 0\;\; 1 \\
 0 \;\;0\;\; 0 \;\;1 \;\;0\;\; 0 & 0\;\; 0\;\; 0\;\; 1\;\; 1\;\; 1 & 1\;\; 0 \;\;1 \;\;0 \\
 0 \;\;0 \;\;0 \;\;0 \;\;1 \;\;0 & 1\;\; 0\;\; 0\;\; 0\;\; 1\;\; 1 & 0\;\; 1\;\; 0\;\; 1 \\
 0\;\; 0\;\; 0\;\; 0\;\; 0 \;\;1 & 1\;\; 1 \;\;0 \;\;0 \;\;0 \;\;1 & 1\;\; 0 \;\;1 \;\;0 \\
 \hline
 0 \;\;0 \;\;0 \;\;0 \;\;0 \;\;0 & 1\;\; 0\;\; 1\;\; 0\;\; 1\;\; 0 & 1\;\; 0\;\; 1\;\; 0 \\
 0 \;\;0\;\; 0 \;\;0 \;\;0 \;\;0 & 0\;\; 1\;\; 0\;\; 1\;\; 0\;\; 1 & 0\;\; 1\;\; 0\;\; 1
\end{array}
\right).
$$
We observe that $G_3 \times H_3^T = 0$, as required.
}
\hfill$\Box$
\end{example}

\begin{remark}
{\rm
It should be noted that Theorem \ref{theorem 1} is valid not only for POT and rPOT ordering but also for any ordering.
We demonstrate Theorem \ref{theorem 1} to the term over position (TOP) ordering \cite{Heegard} on $\mathbb{F}_q [t]^m $
defined by $ t^l e_i  >_\mathrm{TOP} t^k e_j$ if $l>k$, or $l=k$ and $i<j$.
The reduced TOP Gr\"obner basis $\left\{ g_1', g_2', g_3' \right\}$ of $\overline{\mathcal{C}}_3$ turns into
$$
  \left[ \begin{array}{cc}
g_1'\\g_2'\\g_3'
  \end{array}\right]=
  \left[ \begin{array}{ccc}
    1+t, & t+t^2+t^3, & 1+t^2  \\
    1, & 1+t+t^2, & t+t^3   \\
    1+t+t^2, & 0, &  0
\end{array} \right].
$$
It is easy to check that $\left \langle g_i', h_j \right \rangle = 0$ for all $i,j$.
}
\hfill$\Box$
\end{remark}

\section{Estimation of algorithms}\label{algorithms}
In this section, we estimate the computational complexity of two algorithms and compare the one with the other.
We represent the numbers of additions, subtractions, multiplications, and divisions in $\mathbb{F}_q$ as the coefficients of $\kappa$, $\lambda$, $\mu$, and $\nu$, respectively.
We aim for an asymptotic estimation; we denote $f\sim g$ if $f/g$ tends to 1 as the variable tends to $\infty$.
Once we obtain $f\sim g$, then it follows that the usual notation $f=\mathcal{O}(g)$, which means $f\le cg$ for some constant $c>0$.

First, we describe it with respect to the Gaussian elimination. Although it is well-known that the complexity is $\mathcal{O}(n^3)$, we calculate the order up to constant factor.
We can assume that a given $(n-k)\times n$ parity check matrix $H$ is transformed into $[I|A]$ with permutation $\tau=1$ since the permutation costs no finite-field operation.
Without loss of generality, we can assume that the $(1,1)$ component of $H$ is non-zero.
Then, dividing the other component of the first row by this value takes $(n-1)\nu$.
Moreover, subtracting the multiple of the $(i,1)$ component and the first row for $2\le i\le n-k$ takes $(n-k-1)(n-1)(\lambda+\mu)$.
Summing up these manipulations for $n-k$ columns, we obtain
$$
\sum_{i=n-1}^{k+1}\{i\nu+(i-k)i(\lambda+\mu)
+(n-i)k(\lambda+\mu)\}
$$
where the last term $(n-i)k(\lambda+\mu)$ comes from the back substitution.
We ignore the first term $i\nu$ since it contributes square order $\frac12(n+k)(n-k-1)$ of $n$.
Then, we obtain $(\nu+\lambda+\mu)$ times
$$
\frac13(n-k-1)(n-k-\frac12)(n-k)+k(n-k)^2,
$$
which is asymptotically $\frac13(n-k)^3+k(n-k)^2$.

Next, we describe the computational complexity with respect to the Buchberger's algorithm for a given $k\times n$ generator matrix $G$ to obtain a POT Gr\"obner basis.
The estimation is similar to the above; now the algorithm is based on polynomial gcd computation.
Without loss of generality, we can assume that the $(1,l_1)$ component of $G$ is not zero.
Then, dividing the other component of the first row by this value, and moreover, subtracting the multiple of the $(i,l_1)$ component and the first row for $2\le i\le k$ takes $(n-1)\nu+(k-1)(n-1)(\lambda+\mu)$.
The second stage of these manipulations takes $(n-2)\nu+k(n-2)(\lambda+\mu)$ since the first row has a polynomial whose degree is greater than that of the other rows.
Summing up these manipulations for the first orbit, we obtain
$$
\sum_{j=d_1}^{l_1-1}\left[\left(j+\sum_{i=2}^{m}l_i\right)\{\nu+k(\lambda+\mu)\}\right]-(n-1)(\lambda+\mu),
$$
where we denote $d_i:=\deg g_{ii}$ and the last term, which is ignored because of less contribution, comes from the special situation stated above at the first row.
Furthermore, simplifying this and summing up for all orbits, we obtain $(\nu+\lambda+\mu)$ times
\begin{equation}\label{forward}
\sum_{j=1}^{m}(l_j-d_j)\!\left(\frac{l_j-1+d_j}{2}+\!\!\sum_{i=j+1}^{m}l_i\right)\!(k-j+1),
\end{equation}
which is bounded by $nk^2$, since the second bracket $\le n$.

It is necessary that we estimate the reducing computation of Gr\"obner basis, which corresponds to the back substitution of polynomial matrix.
For POT Gr\"obner basis, we must start from reducing $g_{12}$ and not $g_{im}$.
The length of $g_{12}$ is $l_2$ in the worst case, and we have to eliminate $l_2-d_2$ values.
Thus, the reduction of $g_{12}$ takes $(l_2-d_2)\left(d_2+\sum_{i=3}^{m}l_i\right)(\lambda+\mu)$.
Summing up for all $g_{ij}$ $(i<j)$, we obtain $(\lambda+\mu)$ times
\begin{equation}\label{back}
\sum_{j=2}^{m}(l_j-d_j)\left(d_j+\sum_{i=j+1}^{m}l_i\right)(j-1),
\end{equation}
which is bounded by $mnk$.

For the total complexity of Buchberger's algorithm to obtain the reduced Gr\"obner basis, we must add \eqref{back} to \eqref{forward}.
Since we can bound \eqref{back} by the summation of $j$ from $1$, the last bracket of \eqref{forward} is changed into $k$.
Then, we observe that the total complexity is still $nk^2$.

The final stage of estimation is to calculate the number of operations required for computing the polynomial matrix $A$ from the polynomial matrix $(h_{ij})$ by \eqref{successive}.
It should be noted that the multiplication of two polynomial $a(t)$ and $b(t)$ requires $(\deg a\deg b)\kappa + (1+\deg a)(1+\deg b)\mu$ operations, and that the division of $a(t)$ by $b(t)$ requires $\deg b(\deg a - \deg b)(\lambda + \mu)$ operations.
We denote $\epsilon_{i} := \deg h_{ii}$;
then we have
$$
\deg a_{ii}=l_i - \epsilon_{i},\;\;
\sum\limits_{i = 1}^{m} {\epsilon_{i}} = k,\;\;
\sum\limits_{i = 1}^{m} {(l_i - \epsilon_{i})} = n-k.
$$
From \eqref{successive}, we see that the computation of $a_{ij}$ $(j<i)$ is separated into two steps: $\sum\limits_{\delta = j+1}^{i} {a_{i\delta}h_{\delta j}}$ and its division by $h_{jj}$.
Since $\deg a_{i\delta}<l_i - \epsilon_{i}$ and $\deg h_{\delta j} < \epsilon_j$, the complexity of the first step is bounded by
\begin{equation}\label{first step}
\begin{split}
\sum_{\delta=j+1}^{i}&\Big[(l_i - \epsilon_{i}-1)(\epsilon_j-1)\kappa +(l_i - \epsilon_{i})\epsilon_j\mu\Big]\\
&+(i-j-1)(l_i - \epsilon_i + \epsilon_j)\kappa.
\end{split}
\end{equation}
Since we can check by direct calculation that the coefficient of $\kappa$ in \eqref{first step} is bounded by $(i-j)(l_i-\epsilon_i)\epsilon_j$, thus \eqref{first step} is bounded by $(i-j)(l_i-\epsilon_i)\epsilon_j(\kappa+\mu)$.
The second step requires $(l_i - \epsilon_i)\epsilon_j(\lambda+\mu)$.
Hence, the complexity of computing $a_{ij}$ is bounded by $(i-j+1)(l_i - \epsilon_i)\epsilon_j(\kappa + \lambda + \mu)$.
On the other hand, we see from \eqref{successive} that the complexity of computing $a_{ii}$ is bounded by $(l_i - \epsilon_i)\epsilon_i(\lambda + \mu)$, which is viewed as the case of $i=j$ for $(i-j+1)(l_i - \epsilon_i)\epsilon_j(\lambda + \mu)$.
Summing up these results, we obtain $(\kappa + \lambda + \mu)$ times
$$
\sum\limits_{i = 1}^{m} \sum\limits_{j = 1}^{i} (i-j+1)(l_i - \epsilon_{i})\epsilon_{j}
\le mk(n-k).
$$
Therefore, the complexity of computing polynomial matrix $A$ is estimated as $mk(n-k)$.

Thus, we obtain estimation formulae.
\begin{align*}
{\mbox{Echelon canonical}\atop\mbox{form algorithm}}:&\quad\frac13(n-k)^3+k(n-k)^2+nk^2\\
\mbox{Transpose algorithm}:&\quad n(n-k)^2+mnk+mk(n-k)
\end{align*}
We can observe that both algorithms have the same rough order $\mathcal{O}(n^3)$ of computational complexity.

\begin{figure}[t]
\centering
  \resizebox{10.8cm}{!}{\includegraphics{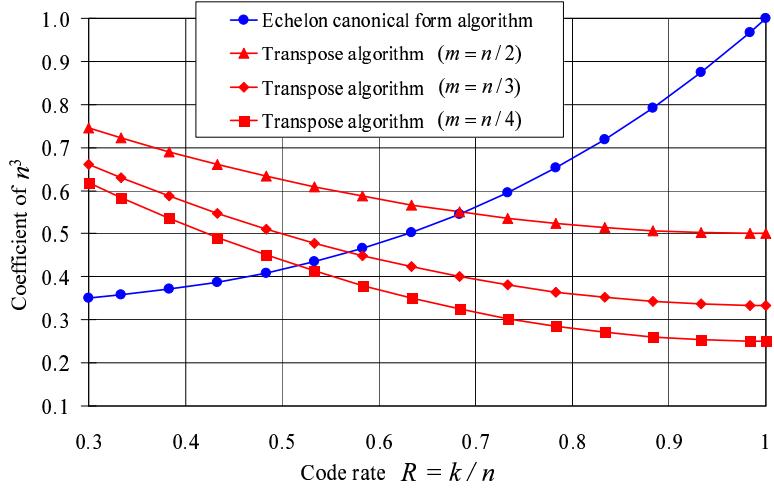}}
\vspace{-4mm}
\caption{Coefficient of $n^3$ in the estimation formulae.\label{graph}}
\end{figure}

For the comparison of two algorithms, we assume $m=\frac12n,\frac13n,\frac14n$ to eliminate $m$ in the estimation formula for transpose algorithm.
Since there exist special GQC codes that satisfy $m=n-1$, these assumptions are not valid for general GQC codes.
However, for effective GQC codes, these assumptions are reasonable; actually, FG LDPC codes have further less number of orbits than these assumptions (cf. Table \ref{comparison} in the next section).
Figure \ref{graph} is the comparison between the coefficients of $n^3$ in the above estimation formulae under the assumptions, where the curves near 1 represent limits.
Thus, we can conclude that, for the effective high-rate GQC codes, the computational complexity of the transpose algorithm is far less than that of the echelon canonical form algorithm.

\section{Estimation of the circuit}\label{circuit}
In the previous sections, we have proposed two algorithms to calculate the reduced Gr\"obner basis of the form \eqref{polynomial matrix} that generates $m$-orbit GQC code $\mathcal{C}$.
In \cite{Chen}, Chen \emph{et al.} have developed a serial-in serial-out hardware architecture to encode information symbols systematically with POT Gr\"obner basis as an application of results in Heegard \emph{et al.} \cite{Heegard}.
The architecture generalizes classical encoder of cyclic codes and consists of division circuits by $g_{ii}(t)$ and multiplication circuits with $g_{ij}(t)$ $(i<j)$.

We quote the estimation of their hardware complexity from \cite{Chen}.
The total numbers of finite-field adder elements $A_m$ and memory elements (shift registers) $D_m$ are given as follows:
\begin{align*}\label{quote}
A_m &\le \sum_{i=1}^m\deg g_{ii}+
\sum_{i=1}^{m-1}\sum_{j=i+1}^m(\deg g_{ij}+1)\\
&\le (n-k)+\sum_{i=1}^{m-1}(m-i)\deg g_{ii}
\le m(n-k),\\
D_m &\le \sum_{i=1}^m\deg g_{ii}+
\sum_{i=1}^{m-1}\sum_{j=i+1}^m{\deg g_{ij}}+
\sum_{i=1}^{m-1} (\delta_i+1)
\le  m(n-k) + k,
\end{align*}
where $\delta_i := \max \left({k_1-2, k_2-2, \cdots, k_{i-1}-2, k_i-1}\right)$, and $k_i:=l_i-\deg g_{ii}$.
We can conclude that the hardware complexity for GQC codes is nearly proportional to the code length since $m$ is small compared to $n$.

\begin{figure*}[t]
\begin{center}
  \includegraphics[scale=0.6]{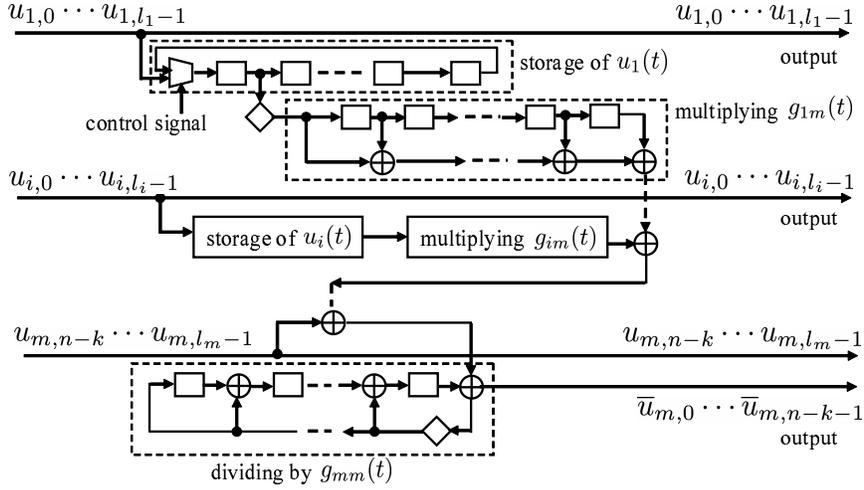}
\vspace{-0.5em}
  \caption{Serial-in serial-out architecture for type-II FG (EG and PG) LDPC codes. Input is information $\{u_{i,j}\}$, and output is redundant parity bits $\{\overline u_{m,j} \}$. The control signal is used to switch for feedback shift registers after entering $u_{1,0},\cdots,u_{1,l_1-1}$.\label{hardware}}
\end{center}
\end{figure*}

For more practical estimation, we focus on the finite geometry (FG) LDPC codes \cite{Kou}\cite{Lin}\cite{Peterson} as an important class of GQC codes.
There are two types of FG LDPC codes: type-I and type-II.
Type-I FG LDPC codes are defined by the parity check matrix composed of incidence vectors (as rows) of lines and points in finite geometries (Euclidean geometry (EG) and projective geometry (PG)) and are cyclic codes.
Type-II FG LDPC codes are defined by the transposed parity check matrix of type-I and are not cyclic but GQC codes.
Therefore, we concentrate on type-II FG LDPC codes.
We quote the required properties of this type of codes from \cite{Tang}.
We denote $n'$ and $k'$ as the corresponding values of type-I codes.
\begin{enumerate}
\item $l_1 \le l_2=\cdots=l_m$ (Actually, it becomes the equality for EG codes.)
\item $g_{11}=\cdots=g_{m-1,m-1}=1$ and $\deg g_{mm}=n-k$
\item $(n-k) = (n' - k') < n'=l_m$
\end{enumerate}
The last two properties follow easily from the fact that the dual codes of FG LDPC codes are the one-generator GQC codes, which corresponds to the case of $l=1$ in [\ref{Tang}, eq.(23)].
Therefore, the reduced POT Gr\"obner basis $\mathcal{G}=\{g_1,\cdots,g_m\}$ of type-II FG LDPC codes
must be in the following form:
$$
\left [
\begin{array}{c}
  g_1 \\
  g_2 \\
  \vdots  \\
  g_{m-1} \\
  g_m
\end{array}
\right ]
 =
\left [
\begin{array}{ccccc}
   1  & 0     & \cdots & 0     & g_{1m}(t)\\
   0  & 1     & \ddots & \vdots& g_{2m}(t)\\
\vdots& \ddots& \ddots & 0     & \vdots \\
\vdots&  &    \ddots   & 1     & g_{m-1,m}(t)\\
   0  & \cdots& \cdots& 0     & g_{mm}(t)
\end{array}
\right],
$$
where $\deg g_{im} < \deg g_{mm}=n-k$.
The information block $u$ is represented as the vector $u=(u_1(t),\cdots,u_m(t))$, where
$$
 u_i (t) = \left\{ \begin{array}{cl}
\sum\limits_{j = 0}^{l_i  - 1} u_{i,j} t^j &
i = 1,\cdots,m - 1,  \\
\sum\limits_{j = n-k}^{l_m  - 1} u_{m,j} t^j & i = m.
\end{array} \right.
$$
The parity block $\overline u=(0,0,\cdots,0,\overline u_m(t))$,
where $\overline u_m(t)= \sum\limits_{j = 0}^{n-k-1} {\overline u_{m,j}} t^j $, is the remainder of $u$ with respect to the reduced Gr\"obner basis $\mathcal{G}$.
The corresponding codeword is the result of subtracted vector $u-\overline u$.
This is received at the output of architecture in Figure \ref{hardware}, serial-in serial-out architecture for FG LDPC codes.
The element $\bigoplus$ represents an adder (exclusive-OR element) and the rectangle represents a memory element (a shift register).
The two remaining building elements correspond to multiplexer and gate elements.
The gate element is a switch control with two status---open and close.
The multiplexer element is signal choice control that selects signal either from input or from the feedback of shift registers.

Then, the total number $A_m$ of adder elements for FG codes satisfies the following inequality:
$$
   A_m\le m(n-k)\le ml_m\le2n.
$$
Moreover, the total number $D_m$ of required memory elements satisfies
$$
D_m\le ml_m + \sum_{i = 1}^{m - 1}l_i= (m-1)l_m+\sum_{i = 1}^m l_i\le 2n.
$$
Thus, we have proved that the hardware complexity of FG LDPC codes is $\mathcal{O}(n)$ order.

For FG LDPC codes made from 3-dimensional EG and PG over the finite field $\mathbb{F}_{2^s}$, where $s = 1,2,3$, we summarize computational results in Table \ref{comparison}.
The last two columns of Table \ref{comparison} are the numbers of adder and memory elements, respectively.
We see that the actual numbers of elements are less than the above estimation.

\begin{table}[t]
\centering
\caption{Hardware complexity for several 3-dimensional type-II EG and PG LDPC codes. The first three rows evaluate type-II EG LDPC codes. The others evaluate type-II PG LDPC codes.\label{comparison}}
\vspace{2mm}
\begin{tabular}{c|c|c|c|c|c|c}
$s$ & $n$ & $k$ & $n-k$ & $m$ &  adder & memory\\ [0.5ex]
\hline
1 & 21      & 15      & 6     & 3  & 12   &  26   \\
2 & 315     & 265     & 50    & 5  & 76   &  328  \\
3 & 4599    & 4227    & 372   & 9  & 1681 &  5769 \\
\hline
1 & 35      & 24      & 11    & 3  & 16   &  36 \\
2 & 357     & 296     & 61    & 5  & 138  &  438 \\
3 & 4745    & 4344    & 401   & 9  & 1846 &  6396 \\
\hline
\end{tabular}
\end{table}

\section{Conclusions}\label{conclusion}
One contribution of this paper is to provide algorithms of computing Gr\"obner basis for efficient systematic encoder of GQC codes.
Our algorithms are applicable to not only binary GQC LDPC codes but also non-binary GQC LDPC codes and linear codes with nontrivial automorphism groups.
Although the computation of Gr\"obner basis is required only once at the construction of encoder differently from decoding algorithm, our algorithms are still useful;
for example, both algorithms can search effective codes rapidly in the polynomial (third power) order of code-length.
For high-rate codes, we have shown that the algorithm applying transpose formula is faster than the echelon canonical form algorithm.
It is expected that GQC LDPC codes improve the decoding performance of QC LDPC codes and make it close to that of the random LDPC codes.
Another contribution of this paper is to demonstrate that the hardware complexity of the serial-in serial-out systematic encoder is the linear order of code-length for FG codes and FG LDPC codes.
By exploiting the structure of GQC codes, we believe that many new and optimum codes are constructed, and our results in systematic encoding might become a key step to practical implementation.

\section*{Acknowledgment}
This work was partly supported
by the Grant-in-Aid for Young Scientists (B, research project 19760269) by the Ministry of Education, Culture, Sports, Science and Technology (MEXT), the Academic Frontier Center by MEXT for ``Future Data Storage Materials Research Project,'' and a research grant from SRC (Storage Research Consortium).

{\bf Vo Tam Van}
received the B.S. and M.S. degrees from the Department of Information Technology, HCM University of Sciences, Vietnam in 2001, 2005, respectively. From 2002 to 2007, he was a lecturer in the Department of Information Technology, HCM University of Sciences, Vietnam. Now, he is a second-year Ph.D. student in the Department of Electronics and Information Science, Toyota Technological Institute, Japan. His research interests include coding theory, algebraic codes, and LDPC codes.
\vspace{1em}

{\bf Hajime Matsui}
received the B.S. degree in 1994 from the Department of Mathematics, Shizuoka University, Japan, and the M.S. degree in 1996 from the Graduate School of Science and Technology, Niigata University, Japan, and the Ph.D. degree in 1999 from the Graduate School of Mathematics, Nagoya University, Japan. From 1999 to 2002, he was a Post-Doctorate Fellow in the Department of Electronics and Information Science, Toyota Technological Institute, Japan. From 2002 to 2006, he was a Research Associate there. Since 2006, he has been working as an Associate Professor there. His research interests include number theory, coding theory, error-correcting codes, and encoding/decoding algorithms. He is a member of SITA and IEEE.
\vspace{1em}

{\bf Seiichi Mita}
received the B.S. degree, the M.S. degree and the Ph.D. degree in electrical engineering from Kyoto University in 1969, 1971, 1989 respectively. He studied at Hitachi Central Research Laboratory, Kokubunji, Japan from 1971 to 1991 on signal processing and coding methods for digital video recording equipment for broadcast use and home use. He moved to Data Storage \& Retrieval Systems Division, Hitachi, Ltd. in 1991. He developed channel coding methods and these LSI chips for magnetic disk drives. Now, he is a professor of Toyota Technological Institute in Nagoya. He is a member of the Institute of Electronics, Information and Communication Engineers and the Institute of Image Information and Television Engineers in Japan. He is also a member of IEEE, Magnetic Society. He received the best paper awards of IEEE Consumer Electronics Society in 1986 and the best paper awards of the Institute of Television Engineers in Japan in 1987.


\begin{thebibliography}{99}

\bibitem{Andrews}
K. Andrews, S. Dolinar, J. Thorpe,
``Encoders for block-circulant LDPC codes,''
Proc. International Symposium on Information Theory and Its applications, Adelaide, Australia, pp.2300-2304, Sep. 2005.

\bibitem{Becker}
T. Becker, V. Weispfenning,
``Gr\"obner bases,''
New York: Springer Publishers, 1992.

\bibitem{Chen}
J.-P. Chen, C.-C. Lu,
``A serial-in serial-out hardware architecture for systematic encoding of Hermitian codes via Gr\"obner bases,''
IEEE Trans. Comm., vol.52, no.8, pp.1322-1331, Aug. 2004.

\bibitem{Cox}
D. Cox, J. Little, D. O'Shea,
``Ideals, Varieties, and Algorithms: An introduction to computational algebraic geometry and commutative algebra,''
2nd ed. New York: Springer Publishers, 1997.

\bibitem{Chung}
S.Y. Chung, G.D. Forney, T.J. Richardson, R.L. Urbanke,
``On the design of low density parity check codes within 0.0045dB of the Shannon limit,''
IEEE Comm. Lett. , vol.5, no.2, pp.58-60, Feb. 2001.

\bibitem{Fan}
J.L. Fan,
``Array codes as low density parity check codes,''
Proc. 2nd International Symposium on Turbo Codes and Related Topics,
Brest, France, pp.543-546, Sep. 2000.

\bibitem{Fossorier}
M.P.C. Fossorier,
``Quasi-cyclic low density parity check codes from circulant permutation matrices,''
IEEE Trans. Inf. Theory, vol.50, no.8, pp.1788-1793, Aug. 2004.

\bibitem{Fujita}
H. Fujita and K. Sakaniwa,
``Some classes of quasi-cyclic LDPC codes: Properties and efficient encoding method,''
IEICE Trans. Fundamentals, vol.E88-A, no.12, pp.3627-3635, Dec. 2005

\bibitem{Gallager}
R. G. Gallager,
``Low density parity check codes,''
IRE Trans. Inf. Theory, vol.IT-8, pp.21-28, Jan. 1962.

\bibitem{Heegard}
C. Heegard, J. Little, K. Saints,
``Systematic encoding via Gr\"obner bases for a class of algebraic geometric Goppa codes,''
IEEE Trans. Inf. Theory, vol.41, no.6, pp.1752-1761, Nov. 1995.

\bibitem{Kaji}
Y. Kaji,
``Encoding LDPC codes using the triangular factorization,''
IEICE Trans. Fundamentals, vol.E89-A, no.10, pp.2510-2518, Oct. 2006

\bibitem{Kamiya}
N. Kamiya, E. Sasaki,
``Design and implementation of high-rate QC-LDPC codes,''
(in Japanese)
Proc. 2006 SITA, Hakodate, Hokkaido, Japan, pp.545-548, Nov. 2006.

\bibitem{Kou}
Y. Kou, S. Lin, M.P.C. Fossorier,
``Low density parity check codes based on finite geometries: A rediscovery and new results,''
IEEE Trans. Inf. Theory, vol.47, no.7, pp.2711-2761, Nov. 2001.

\bibitem{Lally}
K. Lally, P. Fitzpatrick,
``Algebraic structure of quasi-cyclic codes,''
Discrete Applied Mathematics, 111, pp.157-175, Jul. 2001.

\bibitem{Lin}
S. Lin, D.J. Costello,
``Error Control Coding: Fundamentals and Applications,''
2nd ed. Englewood Cliffs, NJ: Prentice-Hall, 2004.

\bibitem{Little}
J. Little, K. Saints, C. Heegard,
``On the structure of Hermitian codes,''
Journal of Pure and Applied Algebra 121, pp.293-314, Oct. 1997.

\bibitem{Little 2007}
J. Little,
``Automorphisms and encoding of AG and order domain codes,''
to appear in volume from D1 Workshop on applications of Gr\"obner bases in coding theory and cryptography, RISC-Linz, 2007.

\bibitem{Mackay}
D.J.C. MacKay,
``Good error-correcting codes based on very sparse matrices,''
IEEE Trans. Inf. Theory, vol.IT-45, no.2, pp.399-431, Mar. 1999.

\bibitem{MacWilliams}
F.J. MacWilliams, N.J.A. Sloane,
``The theory of error correcting codes,''
9th ed. North Holland, 1988.

\bibitem{Maehata}
T. Maehata, M. Onishi,
``A reduced complexity, high throughput LDPC encoder using LU factorization,'' (in Japanese) Proc. the 2008 IEICE general conference, B-5-157, p.543, Mar. 2008.

\bibitem{Peterson}
W.W. Peterson, E.J. Weldon,
``Error correcting codes,''
2nd ed. Cambridge, MA: MIT Press, 1972.

\bibitem{Shokrollahi}
T.J. Richardson, M.A. Shokrollahi, R.L. Urbanke,
``Design of capacity approaching irregular low density parity check codes,''
IEEE Trans. Inf. Theory, vol.IT-47, no.2, pp.619-637, Feb. 2001.

\bibitem{Richardson}
T.J. Richardson, R.L. Urbanke,
``Efficient encoding of low density parity check codes,''
IEEE Trans. Inf. Theory, vol.47, no.2, pp.638-656, Feb. 2001.


\bibitem{Siap}
I. Siap, N. Kulhan,
``The structure of generalized quasi-cyclic codes,''
Applied Mathematics E-Notes, vol.5, pp.24-30, Mar. 2005.

\bibitem{Tang}\label{Tang}
H. Tang, J. Xu, S. Lin, K.A.S. Abdel-Ghaffar,
``Codes on finite geometries,''
IEEE Trans Inf. Theory,
vol.51, no.2, pp.572-596, Feb. 2005.

\bibitem{Tanner}
R.M. Tanner, D. Sridhara, T. Fuja
``A class of group-structured LDPC codes,''
Proc. International Symposium on Communication Theory and Applications, Ambleside, U.K, pp.365-370, Jul. 2001.

\bibitem{Van}
V.T. Van, H. Matsui, S. Mita
``Systematic encoding for finite geometry low density parity check codes based on Gr\"obner bases,''
Proc. 2007 SITA, Kashikojima, Mie, Japan, pp.424-429, Nov. 2007.

\end{thebibliography}
\end{document}